\newcommand{\ket}[1]{\left\vert #1\right\rangle}

\newcommand{\bra}[1]{\left\langle #1\right\vert}

\newcommand{\brkt}[2]{\left\langle #1 \vert #2\right\rangle}

\newcommand{\braket}[3]{\left\langle #1 \right\vert #2\left\vert #3\right\rangle}

\newcommand{\brak}[1]{\left\langle #1\right\rangle}

\newcommand{\etcr}[2]{\left[#1 , #2\right]}

\documentclass[10pt,multicol,onecolumn,amsmath,amssymb]{article}
\usepackage{amsmath, amssymb, graphics, setspace}
\usepackage{times}          
\usepackage{bm}
\usepackage{amsxtra}
\usepackage{amstext}
\usepackage{latexsym}
\usepackage{dsfont}
\usepackage{bbold} 
\usepackage{mathtools}

\usepackage{bm}
\usepackage{amsmath}
\usepackage{amsxtra}
\usepackage{amstext}
\usepackage{amssymb}
\usepackage{latexsym}
\usepackage{dsfont}
\usepackage{bbold} 
\usepackage{mathtools}

\usepackage{url}
\usepackage{array}
\usepackage{booktabs}

\newcommand{\mathsym}[1]{{}}
\newcommand{\unicode}[1]{{}}
\usepackage{multicol}
\usepackage{geometry}
\newcommand{\doublerule}[1][.2pt]{%
  \noindent
  \makebox[0pt][l]{\rule[.7ex]{\linewidth}{#1}}%
  \rule[.3ex]{\linewidth}{#1}}

  \usepackage{hhline}
\usepackage{caption}

{\renewcommand{\thefootnote}{\fnsymbol{footnote}}}
\usepackage{scalerel}
\usepackage{stackengine}
\usepackage{verbatimbox}
\usepackage{boxhandler}
\newlength\dotscale

\reversemarginpar
\usepackage{ytableau}
\usepackage{cite}
\newcommand{\overbar}[1]{\mkern 1.5mu\overline{\mkern-1.5mu#1\mkern-1.5mu}\mkern 1.5mu}

\begin{document}

\begin{center}
{ \textbf{RADIATIVE DECAYS AND THE $\mathbf {SU(6)}$ LIE ALGEBRA}}
\end{center}

\begin{center}

\textbf{{\footnotesize MILTON DEAN SLAUGHTER}$^{*}$}
{\renewcommand{\thefootnote}{\fnsymbol{footnote}}
\setcounter{footnote}{1}
\footnotetext{slaughtm@FIU.Edu,\;\;Slaughts@PhysicsResearch.Net}
\setcounter{footnote}{0}
}
\end{center}

\begin{center}
\textit{{\footnotesize Department of Physics, Florida International University, Miami, Florida 33199, USA}}
\end{center}


\noindent {\footnotesize We present research on radiative decays of vector ($J^{PC}=1^{--}$) to pseudoscalar ($J^{PC}=0^{-+}$) particles ($u$, $d$, $s$, $c$, $b$, $t$ quark system) using broken symmetry techniques in the infinite momentum frame and equal time commutation relations and the $SU(6)$ Lie algebra and conducted without ascribing any specific form to meson quark structure or intra-quark interactions.  We utilize the physical electromagnetic current $j_{em}^{\mu}(0)$ including its singlet $U(1)$ term and focus on the $SU(6)$ $35$-plet. We derive new relations involving the electromagnetic current (including its singlet--proportional to the $SU(6)$ singlet). Remarkably, we find that the electromagnetic current singlet plays an intrinsic role in understanding the physics of radiative decays and that the charged and neutral $\rho$ meson radiative decays into $\pi \, \gamma$ are due entirely to the singlet term in $j_{em}^{\mu}(0)$.  Although there is insufficient radiative decay experimental data available at this time, parametrization of possible predicted values of $\Gamma({{{D}}^{*}}^{0}\rightarrow D^{0}\,\gamma)$ is made.  For conciseness and self-containment, we compute all $SU(6)$ Lie algebra simple roots, positive roots, weights and fundamental weights which allow the construction of all $SU(6)$ representations. We also derive all non-zero $SU(6)$ generator commutators and anti-commutators---useful for further research on grand unified theories.}
\newline





\newpage

\begin{center}
\section*{I. INTRODUCTION}
\end{center}

It is remarkable that in many cases observed particles appear to roughly fit into group-theoretical representation constructs which happen to be special unitary group representations.  While these group-theoretical constructs obviously require that particles belonging to a particular representation all have the same mass, that is not what one observes in the real world\textemdash thus the need for \emph{quark flavor broken symmetry group techniques}.  To date, QCD (quantum chromodynamics) based on Lagrangians (involving the addition of the Higgs field and other terms) invoking spontaneous symmetry breaking is the best theory for describing the real world, although Lattice gauge models are making headway. As is well known, no theory capable of predicting and accommodating physical observations has yet been developed which incorporates the gravitational force. Indeed, although glueballs are predicted to exist in QCD, no \emph{uncontrovertible} candidates have been found.

In this paper, we present research on radiative decays of vector ($J^{PC}=1^{--}$) to pseudoscalar ($J^{PC}=0^{-+}$) particles \cite{ODonnell:1981hgt} which appear to belong\textemdash at least in part\textemdash (especially after application of broken symmetry techniques---infinite momentum frame and asymptotic symmetry is discussed in Section II) to specific flavor $SU_{F}(6)$ \emph{representations}.  The representations of $SU(N)$\textemdash [special (determinant $=$ unity), unitary]\textemdash classical Lie \cite{lie:1888groups} groups are associated with the $SU(N)$ classical, semisimple Lie algebras via linearly independent matrix operators $V_{a}$ [the $V_{a}$ are linear ``charge" generators \cite{GellMann:1964tf}\textemdash and ${{V_{a}}^{\mu}}(x)=\bar{q}^{i}(x){(\lambda_{a}/2)}_{ij}
\gamma^{\mu}q^{j}(x)$ are the corresponding charge density operators ($q$ represents the $u$, $d$%
, $s$, $c$, $b$, $t$ quark system)] which act on the relevant vector space where (bilinear) commutators of the $V_{a}$ are Lie products acting over the real number field.  Each $V_{a}$ is a hermitian 6 x 6  matrix for $N=6$ and there are $6^{2}-1=35$ $V_{a}$, where $a=1,\ldots ,35$.

In addition, we also introduce the singlet $U(1)$ matrix $V_{0}$ which is proportional to the identity matrix and commutes with all other generators and is \emph{explicitly included in the physical electromagnetic current} $j_{em}^{\mu}(0)$.  As we will discover in Section III., \emph{the singlet has an intrinsic role in understanding the physics of radiative decays}.  Indeed, we introduce ``generalized" Gell-Mann matrices [see~Table~1] $\lambda_{a}$ where $V_{a}=\lambda_{a} / 2$ .  We will find that specific combinations of the $V_{a}$ can be ultimately constructed which represent physical "raising" or "lowering" operators and we will label them using $J^{PC}=0^{-+}$ $35$-plet pseudoscalar particle names.  Explicitly\textemdash \emph{in the infinite momentum frame}\textemdash (as we will demonstrate later in this paper)--for example, the \emph{physical} vector charge $V_{K^{0}}$ is $V_{K^{0}}=V_{6}+iV_{7}$ and the \emph{physical} vector charge $V_{\pi^{\pm}}=V_{1}\pm iV_{2}$. The $\lambda_{a}$ satisfy the commutation algebra
$[(\lambda_{a}/2),(\lambda_{b}/2)]=i\sum_{c=1}^{N^{2}-1}f_{abc}(\lambda_{c}/2)$, where the $f_{abc}$ are structure constants [see~Table~2] (we choose $f_{abc}$ to be real and totally antisymmetric under permutations of the indices ${abc}$\textemdash we note that this can be done for $SU(M)$ groups in general). For clarity and conciseness and self-containment, the $SU(6)$ Lie algebra simple roots, positive roots, weights, fundamental weights, non-zero commutators[see~Table~3], and non-zero anti-commutators[see~Table~5] are also determined which allow construction of all $SU(6)$ representations. In Table 4, we also give all non-zero totally symmetric $SU(6)$ tensors $d_{ijk}$ useful in studying quark-gluon scattering and other processes. The $d_{ijk}$ satisfy $d_{ijk}=2 \, \textrm{Tr}(V_{i}\{V_{j},V_{k}\})$.

Unless otherwise specified (or context specified), Lie algebras--(usually denoted by $\mathfrak{su(n)}$)--corresponding to Lie groups $SU(N)$--will just be denoted by $SU(N)$. Thus, $SU(6)$ refers to the compact, analytic, continuous, semisimple Lie algebra for the Lie group $SU(6)$. Cartan \cite{Cartan:1933thesis} denotes $SU(6)$ as $A_{5}$ and $SU(N)$ as $A_{N-1}$.  It is known that the \emph{lowest-dimensional} $SU(N)$ representation is by $N\times N$, traceless matrix generators which we utilize in this paper. Data in this paper is taken from the Particle~Data~Group~ \cite{Tanabashi:2018xqp,Slaughter:charge conjugate states}.

\begin{center}
{\large \textbf{{A. More about the $\mathbf {SU(6)}$ Lie algebra}}}
\end{center}
\emph{The defining bilinear operation\textemdash the commutator [,]\textemdash involving the structure constants and which determines the Lie algebra and a generator scalar product is given by}:
\begin{subequations} \label{eq1}
\begin{align}
&[V_{a},V_{b}]=i\sum_{c=1}^{N^{2}-1}f_{abc}V_{c}\quad\textrm{where    } a,b=1,2,\ldots, N^{2}-1,\label{eq1B}\\
&\text{Tr}\left[V_a\,V_b\right]=\frac{1}{2}\delta _{ab}, \quad \textrm{and   } \label{eq2C}\\
&f_{abc}=-i \,2 \,\text{Tr}\,(\,[V_{a},V_{b}]\,V_{c})\; \label{eq3D}.
\end{align}
\end{subequations}

The structure constants and Lie generators have been constructed so that \emph{they remain the same for $SU(n-m)$}, where $n>m$ ($n$, $m$ are positive integers). In general, commutators of Lie group generators [see~Table~3] are themselves linear combinations of these same generators and the generator algebra is called a Lie algebra \cite{Dynkin1950AMS,Racah:1961sj,behrends1962simple,Lichtenberg:1978pc,Slansky:1981yr,georgi:1999lie,
cahn:1984lie,Hamermesh:1123140,Oneda:1985wf}. If the group has $r \,\textrm {(group order)}\equiv N^{2}-1$ generators, then there are $(1/2)(r-1)r=(1/2)(N^{2}-2)(N^{2}-1)$ possible generator commutation relations. The rank $l$ of a Lie group is equal to the \emph{maximum} number of generators (linear) which \emph{mutually commute}. There also exist $l$ \emph{nonlinear} Casimir operators $C_{i}=\sum_{j=1}^{r}a_{j i}V_{j}^{i+1}$, $i=1,2,\ldots,l$, which commute with \emph{all} of the algebra generators. Following primarily the notation of Lichtenberg \cite{Lichtenberg:1978pc} but also others: \cite{Slansky:1981yr,georgi:1999lie,cahn:1984lie,Hamermesh:1123140,Oneda:1985wf,Oneda:1991wz,
jones:1998groups,Shapiro:2017lectures,Weyl:2016Stanford,Weyl:1927vd,Christoph:2010}\textemdash the \emph{mutually commuting}  generators are conventionally denoted by $H_{i}$  ($i=1,2,\ldots, l\,$), where $H_{i}={H_{i}}^{\dag}$, and $[H_{i},H_{j}]=0$ ($i,j=1,2,\ldots, l\,$).
So $H_{1}=V_{3}= \dfrac{\lambda_{3}}{2}, \, H_{2}=V_{8}= \dfrac{\lambda_{8}}{2}, \, H_{3}=V_{15}= \dfrac{\lambda_{15}}{2}, \, H_{4}=V_{24}= \dfrac{\lambda_{24}}{2}, \, H_{5}=V_{35}= \dfrac{\lambda_{35}}{2}$ and
  $\vec{H}\equiv\left(H_{1}, H_{2}, H_{3}, H_{4}, H_{5} \right)$.  The $l$ (maximal number of \emph{mutually commuting hermitian generators}) $H_{i}$ [\emph{Cartan generators}] are the basis for what is called the \emph{Cartan subalgebra} and constitute a linear space.  The $H_{i}$ are Cartan ($A_{5}$) Generators and the rank of the traceless, semi-simple, compact Lie algebra of the \textit{classical} group $SU(6)$ is $(6-1)=5 =$ number of $H_{i}$'s.

\begin{center}
{\large \textbf{{B. Roots and weights}}}
\end{center}

  Given the generators $V_{a}$, one can construct $A=\sum_{j=1}^{r} a_{j}V_{j}$ and the eigenvalue equation $[A,X]=\rho X$, where $X$ is some linear combination of the $V_{j}$ , then one can derive the secular equation (polynomial of degree $r$) for the $(r-l)$ eigenvalues called \emph{roots} $\rho $, namely $\det(\sum_{i =1}^{r} a_{i}f_{i j k}-\rho\delta_{j k})=0$.  For semisimple Lie groups (which includes $SU(N)$), Cartan has shown that are $r$ independent eigenvectors (even if there exist degenerate roots for $\rho=0$) and the multiplicity of these degenerate roots is equal to the rank $l$.

We define (Cartan-Weyl formalism) $l=5$ generators [the $H_{i}$ mentioned above],  $(r-l)=30$ remaining generators, $E_{\alpha}\equiv V_{\alpha}$ where $\alpha= [\pi ^{\pm},K^{\pm},K^{0},\bar{K}^{0},\bar{D}^{0},D^{0},D^{\pm},D_{s}^{\pm},B^{0},\bar{B}^{0},
B^{\pm},B_{s}^{0},\bar{B}_{s}^{0},B_{c}^{\pm},T^{\pm},T^{0},\bar{T}^{0},T_{s}^{\pm},\\
T_{c}^{0}, \bar{T}_{c}^{0},T_{b}^{\pm}$], six-dimensional basis states (vectors $u_{i}$), the diagonal vector operator $\vec{H}$, and six five\nobreakspace-dimensional \emph{weights} (\emph{convention dependent}) $\vec{m}(i)$ of $SU(6)$:

\begin{subequations}\label{eqbasic2}
\begin{align}
  H_{i}={H_{i}}^{\dag}, \quad [H_{i},H_{j}]=0,\quad \text{Tr} \ [H_{i} \ H_{j}]=\frac{1}{2}\delta_{ij}, \quad \textrm{and} \quad \vec{H}\equiv\left(H_{1}, H_{2}, H_{3}, H_{4}, H_{5} \right),\hspace{.2in}\label{eqbasic2B}\\
  u_{i}=
  \left(
    \begin{array}{c}
      \delta_{i}^{1} \\
      \delta_{i}^{2} \\
      \delta_{i}^{3} \\
      \delta_{i}^{4} \\
      \delta_{i}^{5} \\
      \delta_{i}^{6} \\
    \end{array}
  \right)
   \Rightarrow u_{1}=\ket{u}=\left(
                               \begin{array}{c}
                                 1 \\
                                 0 \\
                                 0 \\
                                 0 \\
                                 0 \\
                                 0 \\
                               \end{array}
                             \right)
   ,u_{2}=\ket{d}=\left(
                               \begin{array}{c}
                                 0 \\
                                 1 \\
                                 0 \\
                                 0 \\
                                 0 \\
                                 0 \\
                               \end{array}
                             \right),\ldots,u_{6}=\ket{t}=\left(
                               \begin{array}{c}\label{eqbasic2C}
                                 0 \\
                                 0 \\
                                 0 \\
                                 0 \\
                                 0 \\
                                 1 \\
                               \end{array}
                             \right),\\
                             \vec{H}u_{i}=\vec{m}(i)u_{i},\quad \quad \vec{m}(i)=\left(m_{1},m_{2}, m_{3},m_{4},m_{5}, m_{6}  \right) ,\label{eqbasic2D}\\
   \vec{m}(1)=\left(H_{1}u,H_{2}u, H_{3}u,H_{4}u,H_{5}u   \right),\quad
   \vec{m}(2)=\left(H_{1}d,H_{2}d, H_{3}d,H_{4}d,H_{5}d   \right),\label{eqbasic2E}\\
   \vec{m}(3)=\left(H_{1}s,H_{2}s, H_{3}s,H_{4}s,H_{5}s   \right),\quad
   \vec{m}(4)=\left(H_{1}c,H_{2}c, H_{3}c,H_{4}c,H_{5}c   \right),\label{eqbasic2F}\\
   \vec{m}(5)=\left(H_{1}b,H_{2}b, H_{3}b,H_{4}b,H_{5}b   \right),\quad
   \vec{m}(6)=\left(H_{1}t,H_{2}t, H_{3}t,H_{4}t,H_{5}t   \right).\label{eqbasic2G}
  \end{align}
\end{subequations}

Using Equations~(\ref{eqbasic2}), Table~1 and  Table~3, we obtain:

\begin{subequations}\label{eqbasic1}
\begin{equation}
 \vec{m}(1)=\left(\frac{1}{2},\frac{1}{2 \sqrt{3}},\frac{1}{2 \sqrt{6}},\frac{1}{2
   \sqrt{10}},\frac{1}{2 \sqrt{15}}\right), \quad
   \vec{m}(2)=\left(-\frac{1}{2},\frac{1}{2 \sqrt{3}},\frac{1}{2 \sqrt{6}},\frac{1}{2
   \sqrt{10}},\frac{1}{2 \sqrt{15}}\right), \\
 \end{equation}
   \begin{equation}
   \vec{m}(3)=\left(0,-\frac{1}{\sqrt{3}},\frac{1}{2 \sqrt{6}},\frac{1}{2
   \sqrt{10}},\frac{1}{2 \sqrt{15}}\right), \quad \vec{m}(4)=\left(0,0,-\frac{3}{2\sqrt{6}},\frac{1}{2 \sqrt{10}},\frac{1}{2
   \sqrt{15}}\right),\\
\end{equation}
\begin{equation}
  \vec{m}(5)=\left(0,0,0,-\frac{2}{\sqrt{10}},\frac{1}{2 \sqrt{15}}\right), \quad
  \vec{m}(6)=\left(0,0,0,0,-\frac{5}{2\sqrt{15}}\right)\; ,
\end{equation}
\begin{equation}
  \vec{m}(i)\cdot\vec{m}(j)=-\frac{1}{2*6}+\frac{1}{2}\delta_{ij}\; ,
\end{equation}
\begin{equation}
  \sum_{i=1}^{i=\,6}\vec{m}(i)=0 \,.
\end{equation}
\end{subequations}

\parbox{3in}{So the $j$-th component of $\vec{m}(k)$ is given by:}
\begin{equation}
   m_{j}(k)=
\begin{dcases}
    [2j(j+1)]^{-1/2} & \quad k<j+1 \quad ,\\
    -j[2j(j+1)]^{-1/2} &  \quad k=j+1 \quad ,\\
    0 &, \quad k>j+1 \quad .\label{eqbasic3}
\end{dcases}
\end{equation}

The eigenvalues $\vec{m}(i)$ form a 5-simplex hexateron \cite{georgi:1999lie, Cox:1973}
and are the \emph{weights of the first fundamental representation} of $SU(6)$ spanning a $l$-dimensional vector \emph{weight-space}.  We use the convention that $\vec{m}$ is higher than
${\vec{m}}^{'}$ if the last component of the vector $\vec{m}-{\vec{m}}^{'}$ is positive\textemdash if that is zero, one moves to the next component and so on. Thus, the
(eigenvalues) $\vec{m}(i)$ in Equation~(\ref{eqbasic1}) are already ordered.  There exist four other fundamental representations of $SU(6)$, however one can construct the entire Lie algebra from the first fundamental representation.


The positive roots $\alpha_{i}$ are:
\begin{equation}\label{root:positive}
\vec{m}(i) - \vec{m}(j)\quad \textrm{for}\quad i < j \quad
(i, j)= 1,\ldots ,6.
\end{equation}
There are 15 positive roots.

We now introduce some helpful new notation:

\begin{equation}\label{eq:roots}
\parbox{.7in}{
\textrm{$\bf{SU(6)}$ \textbf{$\mathbf{6\,x\,6}$ Lie Algebra Generator Matrix where each element is also a} $\mathbf{6\,x\,6}$ \textbf{Matrix}. }
}
\begin{dcases*}
\quad
\left(
 \begin{array}{cccccc}
 V_3 & V_1^2 & V_1^3 & V_1^4 & V_1^5 & V_1^6 \\
 V_2^1 & V_8 & V_2^3 & V_2^4 & V_2^5 & V_2^6 \\
 V_3^1 & V_3^2 & V_{15} & V_3^4 & V_3^5 & V_3^6 \\
 V_4^1 & V_4^2 & V_4^3 & V_{24} & V_4^5 & V_4^6 \\
 V_5^1 & V_5^2 & V_5^3 & V_5^4 & V_{35} & V_5^6 \\
 V_6^1 & V_6^2 & V_6^3 & V_6^4 & V_6^5 & V_0 \\
\end{array}
\right) \equiv
\left(
 \begin{array}{cccccc}
 V_3 & V_{\pi^{+}} & V_{K^+} & V_{\bar{D}^0} & V_{B^+} & V_{\bar{T}^0} \\
 V_{\pi ^-} & V_8 & V_{K^0} & V_{D^-} & V_{B^0} & V_{T^-} \\
 V_{K^-} & V_{\bar{K}^0} & V_{15} & V_{D_{s}^{-}} & V_{B_{s}^{0}} & V_{T_{s}^{-}} \\
 V_{D^0} & V_{D^+} & V_{D_{s}^{+}} & V_{24} & V_{B_{c}^{+}} & V_{\bar{T}_{c}^{0}} \\
 V_{B^-} & V_{\bar{B}^0} & V_{\bar{B}_{s}^{0}} & V_{B_{c}^{-}} & V_{35} & V_{T_{b}^{-}} \\
 V_{T^0} & V_{T^+} & V_{T_{s}^{+}} & V_{T_{c}^{0}} & V_{T_{b}^{+}} & V_0 \\
\end{array}
\right)  \\ \\
 \quad = \left(
 \begin{array}{cccccc}
 V_3 & V_{1}+i\ V_{2} & V_{4}+i\ V_{5} & V_{9}+i\ V_{10} & V_{16}+i\ V_{17} & V_{25}+i\ V_{26} \\
 V_{1}-i\ V_{2} & V_8 & V_{6}+i\ V_{7} & V_{11}+i\ V_{12} & V_{18}+i\ V_{19} & V_{27}+i\ V_{28} \\
 V_{4}-i\ V_{5} & V_{6}-i\ V_{7} & V_{15} & V_{13}+i\ V_{14} & V_{20}+i\ V_{21} & V_{29}+i\ V_{30} \\
 V_{9}-i\ V_{10} & V_{11}-i\ V_{12} & V_{13}-i\ V_{14} & V_{24} & V_{22}+i\ V_{23} & V_{31}+i\ V_{32} \\
 V_{16}-i\ V_{17} & V_{18}-i\ V_{19} & V_{20}-i\ V_{21} & V_{22}-i\ V_{23} & V_{35} & V_{33}+i\ V_{34} \\
 V_{25}-i\ V_{26} & V_{27}-i\ V_{28} & V_{29}-i\ V_{30} & V_{31}-i\ V_{32} & V_{33}-i\ V_{34} & V_0 \\
\end{array}
\right) \stackrel{\bf{\strut(i\neq j)}}{\equiv} V_{i}^{j}   \\ \quad [\textrm{i= row idex, j=column index}]\textrm{ ,}\\ \\
 \qquad \qquad \begin{matrix}
 V_1^1\equiv V_3+\frac{1}{\sqrt{3}}V_{8}+\frac{1}{\sqrt{6}}V_{15}+\frac{1}{\sqrt{10}}V_{24}+\frac{1}{\sqrt{15}}V_{35}, \\
 V_2^2\equiv -V_3+\frac{1}{\sqrt{3}}V_{8}+\frac{1}{\sqrt{6}}V_{15}+\frac{1}{\sqrt{10}}V_{24}+\frac{1}{\sqrt{15}}V_{35}, \\
 V_3^3\equiv -\frac{2}{\sqrt{3}}V_{8}+\frac{1}{\sqrt{6}}V_{15}+\frac{1}{\sqrt{10}}V_{24}+\frac{1}{\sqrt{15}}V_{35},\\
 V_{4}^{4}\equiv \frac{3}{\sqrt{6}}V_{15}+\frac{1}{\sqrt{10}}V_{24}+\frac{1}{\sqrt{15}}V_{35},\\
 V_{5}^{5}\equiv -\frac{4}{\sqrt{10}}V_{24}+\frac{1}{\sqrt{15}}V_{35}\textrm{,}\\
 V_{6}^{6}\equiv \frac{5}{\sqrt{15}}V_{35} \textrm{  ,}\\
 \end{matrix}\\
 \quad \textrm{and}\\
\qquad \qquad
\begin{matrix}
 V_3=\frac{1}{2}(V_1^1-V_2^2),  \\
 V_{8}=\frac{1}{2\sqrt{3}}(V_1^1+V_2^2-2V_3^3), \\
 V_{15}=\frac{1}{2\sqrt{6}}(V_1^1+V_2^2+V_3^3-3V_{4}^{4}), \\
 V_{24}=\frac{1}{2\sqrt{10}}(V_1^1+V_2^2+V_3^3+V_4^4-4V_{5}^{5}),\\
 V_{35}=\frac{1}{2\sqrt{15}}(V_1^1+V_2^2+V_3^3+V_4^4+V_5^5-5V_6^6).
 \\
 \end{matrix}
\end{dcases*}
\end{equation}

We also have:

\begin{equation}\label{eqbasic4}
  \etcr{\vec{H}}{E_{\alpha}}=\etcr{\vec{H}}{V_{\alpha}}=\vec{\rho}\,(\alpha)\;V_{\alpha},
\end{equation}
where the $\vec{\rho}\,(\alpha)\;$ are $l$-dimensional \textrm{\textbf{root vectors}} spanning a $(r-l)=30$ dimensional root-space.
\\

We can extract all roots $\vec{\rho}\,(\alpha)\;$ from the non-zero commutation relations given in Table~3 and using Equation~(\ref{eqbasic4}). \emph{We define a root as positive if its last component is positive}--otherwise it is negative.  In addition, it can be shown that $\vec{\rho}(\alpha^{\dag})=-\vec{\rho}(\alpha)$ so that for instance $\vec{\rho}(\pi^{-})=-\vec{\rho}(\pi^{+})=(-1,0,0,0,0)$ .
Thus, for $V_{\alpha}$ \,where  [$\alpha=  \pi ^{-},\pi ^{+},K^{+},K^{0},K^{-},\bar{K}^{0},\bar{D}^{0},\ldots,B^{+},\ldots,\bar{T}^{0},\ldots , T_{b}^{+}$], utilizing Equations~(\ref{eqbasic2}), Table~1 (defines the $V_{\alpha}$ in terms of the $\lambda$ matrices), and Table~3 (which lists all non-zero commutators) , one can extract the positive root vectors $\vec{\rho}\,(\alpha)$.  One can also obtain the positive roots by using:

\begin{equation}\label{root:positive1}
\textrm{positive roots are given by:}\quad\vec{m}(i) - \vec{m}(j)\quad \textrm{for}\quad i < j \,.
\end{equation}

Either way, we obtain the following positive root listing\textemdash{Note that the positive roots lie to the right of the diagonal of the $\mathbf{6\,x\,6}$ Lie Algebra Generator Matrix in Equation~(\ref{eq:roots}) and have length $1$:

\begin{table}[!ht]
\begin{center}
\begin{tabular}{| c | c | c |}
\toprule[1.5pt]
$\vec{\rho}(\pi^{+})=(1,0,0,0,0)$ & $\vec{\rho}(K^{+})=(\frac{1}{2},\frac{3}{2 \sqrt{3}},0,0,0)$ & $\vec{\rho}(\overline{D}^{0})=(\frac{1}{2},\frac{1}{2 \sqrt{3}},\frac{2}{\sqrt{6}},0,0)$  \\
$\vec{\rho}(B^{+})=(\frac{1}{2},\frac{1}{2 \sqrt{3}},\frac{1}{2 \sqrt{6}},\frac{5}{2 \sqrt{10}},0)$ &$\vec{\rho}(\overline{T}^{0})=(\frac{1}{2},\frac{1}{2 \sqrt{3}},\frac{1}{2 \sqrt{6}},\frac{1}{2 \sqrt{10}},\frac{3}{\sqrt{15}})$   & $\vec{\rho}(K^{0})=(-\frac{1}{2},\frac{3}{2 \sqrt{3}},0,0,0)$    \\
$\vec{\rho}(D^{-})=(-\frac{1}{2},\frac{1}{2 \sqrt{3}},\frac{2}{\sqrt{6}},0,0)$    & $\vec{\rho}(B^{0})=(-\frac{1}{2},\frac{1}{2 \sqrt{3}},\frac{1}{2 \sqrt{6}},\frac{5}{2 \sqrt{10}},0)$  & $\vec{\rho}(T^{-})=(-\frac{1}{2},\frac{1}{2 \sqrt{3}},\frac{1}{2 \sqrt{6}},\frac{1}{2 \sqrt{10}},\frac{3}{\sqrt{15}})$   \\
$\vec{\rho}(D_{s}^{-})=(0,-\frac{1}{\sqrt{3}},\frac{2}{\sqrt{6}},0,0)$   & $\vec{\rho}(B_{s}^{0})=(0,-\frac{1}{\sqrt{3}},\frac{1}{2 \sqrt{6}},\frac{5}{2 \sqrt{10}},0)$    & $\vec{\rho}(T_{s}^{-})=(0,-\frac{1}{\sqrt{3}},\frac{1}{2 \sqrt{6}},\frac{1}{2 \sqrt{10}},\frac{3}{\sqrt{15}})$    \\
$\vec{\rho}(B_{c}^{+})=(0,0,-\frac{3}{2 \sqrt{6}},\frac{5}{2 \sqrt{10}},0)$   &$\vec{\rho}(\overline{T}_{c}^{0})=(0,0,-\frac{3}{2 \sqrt{6}},\frac{1}{2 \sqrt{10}},\frac{3}{\sqrt{15}})$     & $\vec{\rho}(T_{b}^{-})=(0,0,0,-\frac{2}{\sqrt{10}},\frac{3}{\sqrt{15}})$    \\
\bottomrule[1.5pt]
\end{tabular}
\caption*{\textrm{\textbf{List of Positive Root Vectors}}}
\end{center}
\begin{center}
\end{center}
\end{table}

Of the $15$ positive roots, only $l=5$ are linearly independent and complete and are called the simple roots.
The number of simple roots is equal to the rank of the algebra, the number of Cartan generators.

The simple roots are:
\begin{equation}\label{root:simple1}
\vec{\rho}({\alpha_{i}})=\vec{m}(i) - \vec{m}(i+1) \quad \textrm{for}\quad i=1\ldots (6-1)=5 \,.
\end{equation}

\begin{table}[!ht]
\begin{center}
\end{center}
\begin{center}
\begin{tabular}{c  c  c }
\toprule[1.5pt]
$\vec{\rho}(\pi^{+})=(1,0,0,0,0)$  & $\vec{\rho}(K^{0})=(-\frac{1}{2},\frac{3}{2 \sqrt{3}},0,0,0)$            & $\vec{\rho}(D_{s}^{-})=(0,-\frac{1}{\sqrt{3}},\frac{2}{\sqrt{6}},0,0)$ \\
$\vec{\rho}(B_{c}^{+})=(0,0,-\frac{3}{2 \sqrt{6}},\frac{5}{2 \sqrt{10}},0)$ & $\vec{\rho}(T_{b}^{-})=(0,0,0,-\frac{2}{\sqrt{10}},\frac{3}{\sqrt{15}})$ & \\
\bottomrule[1.5pt] 
\end{tabular}
\caption*{\textrm{\textbf{List of Simple Root Vectors}}}
\end{center}
\end{table}

\begin{center}
{\large \textbf{{C. Fundamental weights}}}
\end{center}

The fundamental representation weights $\vec{\mu}_{j}$ of $SU(6)$ are given by:

\begin{equation}\label{root:simple}
  \frac{2 \, \vec{\rho}(\alpha_{i}) \cdot \vec{\mu}_{j}}{{\alpha_{i}}^{2}}=\delta_{ij} \quad \textrm{and}
\end{equation}

\begin{equation}\label{weight:fundamental}
\vec{\mu}_{j}=\sum_{k=1}^{j}\vec{m}(k) \, .
\end{equation}



\begin{table}[!ht]
\begin{center}
\end{center}
\begin{center}
\begin{tabular}{c  c  c }
\toprule[1.5pt]
$\vec{\mu}_{1}=(\frac{1}{2} , \frac{1}{2 \sqrt{3}} , \frac{1}{2 \sqrt{6}} , \frac{1}{2 \sqrt{10}} , \frac{1}{2 \sqrt{15}})$
    &
  $\vec{\mu}_{2}=(0 , \frac{1}{\sqrt{3}} , \frac{1}{\sqrt{6}} , \frac{1}{\sqrt{10}} , \frac{1}{\sqrt{15}})$
              &
$\vec{\mu}_{3}=( 0 , 0 , \frac{3}{2\sqrt{6}} , \frac{3}{2 \sqrt{10}} , \frac{3}{2\sqrt{15}})$ \\
$\vec{\mu}_{4}=( 0 , 0 , 0 , \frac{2}{\sqrt{10}} , \frac{2}{\sqrt{15}})$   & $\vec{\mu}_{5}=( 0 , 0 , 0 , 0 , \frac{5}{2\sqrt{15}})$  &  \\
\bottomrule[1.5pt] 
\end{tabular}
\caption*{\textrm{\textbf{List of $\bf{SU(6)}$ Fundamental Weights}}}
\end{center}
\end{table}

The rank $l$ $\vec{\mu}_{j}$ are the highest (dominant) weights of the rank $l$ fundamental representations and are complete.  Thus, the highest weight $\vec{\mu}$ of \emph{any} irreducible $\bf{SU(6)}$ representation can be written in terms of these $\vec{\mu}_{j}$. Indeed, we have:

\begin{equation}\label{eq:dynkin1}
  \frac{2 \, \vec{\rho}(\alpha_{j}) \cdot \vec{\mu}}{{\alpha_{j}}^{2}}=l_{j} \, ,
\end{equation}
\begin{equation}\label{eq:dynkin}
 \vec{\mu}=\sum_{j=1}^{l}l_{j}\;\vec{\mu}_{j} \, ,
\end{equation}

where the $l_{j}$ are Dynkin coefficients and are non-negative integers. The $\bf{SU(6)}$ $\mathbf{35}$\textendash \textbf{plet} is denoted by $(1,0,0,0,1)$.  The \emph{standard} Young tableau can be constructed by noting that the k-th Dynkin label is the number of tableau columns with k boxes.  Thus, for instance:

\mbox{
$(1,1,0,0,0)\, \sim$  \ytableausetup{centertableaux}
\begin{ytableau}
\phantom{1} &   \\
\phantom{1}
\end{ytableau} \, \, .
}
In general (see Ref.~\cite{Lichtenberg:1978pc}\,), a tableau is specified by $\vec{p}=(p_{1},p_{2},p_{3},p_{4},p_{5})$ (here the $p_{j}$ = positive integers $=l_{j}$ [$j=1,\ldots,l$] in Equation~(\ref{eq:dynkin})) and the \emph{i}--th \emph{fundamental representation} is given by:
\begin{equation}\label{eq:pfundamental}
  p_{i}=1 \qquad p_{j}=0 \qquad \textrm{where  }i\neq j \quad \textrm{for}\quad i,j=1\ldots (6-1)=5 \,.
\end{equation}

Thus, \mbox{
$(1,0,0,0,0)\, \sim$  \ytableausetup{centertableaux}
\begin{ytableau}
\phantom{1}
\end{ytableau} \,$\sim$ \, \textrm{first fundamental representation of } ${\bf SU(6)}$ .
}
We also note that for an irreducible representation (irrep) with $\vec{p}=(p_{1},p_{2},p_{3},p_{4},p_{5})$ the highest weight is given by :
\begin{equation}\label{eq:dominantweight}
 \vec{\mu}_{dominant\,weight}=\sum_{i=1}^{l}\;p_{i}\vec{\mu}_{i} \, .
\end{equation}

So the highest weight for the irrep $(1,1,0,0,0)$ is $(\frac{1}{2},\frac{3}{2\sqrt{3}},\frac{3}{2 \sqrt{6}},\frac{3}{2 \sqrt{10}},\frac{3}{2 \sqrt{15}})$, whereas the highest weight for the irrep $(1,0,0,0,1)$ is $(\frac{1}{2},\frac{1}{2 \sqrt{3}},\frac{1}{2 \sqrt{6}},\frac{1}{2 \sqrt{10}},\frac{3}{\sqrt{15}})$.

The dimension $D_{6}$ of a particular representation with $\vec{p}=(p_{1},p_{2},p_{3},p_{4},p_{5})$ is given by the following equation (See~Ref.~\cite{Lichtenberg:1978pc}\,):

\begin{align}\label{eq:SU6Dimension}
 D_{6}(p_{1},p_{2},p_{3},p_{4},p_{5})&=\frac{1}{2!3!4!5!}(p_{1}+1)(p_{1}+p_{2}+2)\nonumber\\
 \cdot \, &(p_{1}+p_{2}+p_{3}+3)\nonumber\\
 \cdot \, &(p_{2}+1)(p_{2}+p_{3}+2)(p_{3}+1)\nonumber\\
 \cdot \, &(p_{4}+1)(p_{4}+p_{3}+2)(p_{4}+p_{3}+p_{2}+3)(p_{4}+p_{3}+p_{2}+p_{1}+4)\nonumber\\
 \cdot\,&(p_{5}+1)(p_{5}+p_{4}+2)(p_{5}+p_{4}+p_{3}+3)(p_{5}+p_{4}+p_{3}+p_{2}+4)\nonumber\\
 \cdot\,&(p_{5}+p_{4}+p_{3}+p_{2}+p_{1}+5).
\end{align}

\section*{II.  EQUAL TIME COMMUTATION AND ANTI-COMMUTATION \newline RELATIONS AND INFINITE MOMENTUM FRAME ASYMPTOTIC SYMMETRY}

\begin{equation}\label{eqbasic8}
\parbox{3in}{
$\bf{SU(6)}$ $\mathbf{35}$\textendash \textbf{plet} $\mathbf{q\bar{q}}$ \, \textbf{Representation Matrix}
}
\begin{dcases*}
\bordermatrix{&\bar{u}&\bar{d}&\bar{s}&\bar{c}&\bar{b}&\bar{t}\cr
u &   u \bar{u} & u \bar{d} & u \bar{s} & u \bar{c} & u \bar{b} & u \bar{t} \cr
d &   d \bar{u} & d \bar{d} & d \bar{s} & d \bar{c} & d \bar{b} & d \bar{t} \cr
s &   s \bar{u} & s \bar{d} & s \bar{s} & s \bar{c} & s \bar{b} & s \bar{t} \cr
c &   c \bar{u} & c \bar{d} & c \bar{s} & c \bar{c} & c \bar{b} & c \bar{t} \cr
b &   b \bar{u} & b \bar{d} & b \bar{s} & b \bar{c} & b \bar{b} & b \bar{t} \cr
t &   t \bar{u} & t \bar{d} & t \bar{s} & t \bar{c} & t \bar{b} & t \bar{t} \cr}
\end{dcases*}
\end{equation}

\begin{equation}\label{eqbasic9}
\parbox{4.5in}{
\textrm{$\bf{SU(6)}$} \textbf{Normalized, Orthogonal, and Traditional Zero Weight} 
 \textbf{Particle Representation States Constructed Using Diagonal $\bf{SU(6)}$ Group Quark Matrix Elements}.
}
\begin{dcases*}
  \begin{array}{l}
    \left|\eta _3 \right\rangle =\left| \frac{u \bar{u}-d \bar{d}}{\sqrt{2}}\right\rangle = \left| \pi^0 \right\rangle
    \\
    \left| \eta _8\right\rangle =\left| \frac{u \bar{u}+d \bar{d}-2 s \bar{s}}{\sqrt{6}}\right\rangle \sim \left| \eta \right\rangle
    \\
    \left| \eta _{15}\right\rangle =\left| \frac{u \bar{u}+d \bar{d}+s \bar{s}-3 c \bar{c}}{2 \sqrt{3}}\right\rangle
    \\
    \left| \eta _{24}\right\rangle =\left| \frac{u \bar{u}+d \bar{d}+s \bar{s}+c \bar{c}-4 b \bar{b}}{2 \sqrt{5}}\right\rangle
    \\
    \left| \eta _{35}\right\rangle =\left| \frac{u \bar{u}+d \bar{d}+s \bar{s}+c \bar{c}+b \bar{b}-5 t \bar{t}}{\sqrt{30}}\right\rangle
    \\
    \left| \eta _0\right\rangle =\left| \frac{u \bar{u}+d \bar{d}+s \bar{s}+c \bar{c}+b \bar{b}+t \bar{t}}{\sqrt{6}}\right\rangle
  \end{array}
  \end{dcases*}
\end{equation}

In Ref.~\cite{Slaughter:2011xs}, using infinite-momentum frame broken asymptotic symmetry, we calculated the magnetic moments of the \emph{physical} on-mass shell $J^{P}={3/2}^{+}$ ground-state decuplet baryons without ascribing any specific form to their quark structure or intra-quark interactions by using \emph{equal-time commutation relations }(ETCRs) which involve at most one current density, thus, avoiding problems associated with Schwinger terms.  Here, the ETCRs involve the vector charge generators (the $V_{\alpha }$) of the symmetry groups of QCD. They are valid even though these symmetries are broken
\cite{Oneda:1985wf,Oneda:1991wz,Slaughter:2011xs,Slaughter:2008zd,Oneda:1989jr,Oneda:1989ik,
Slaughter:1988hx,Oneda:1979mp,alfaro63,GellMann:1964tf,adler,Weisberger} and even when the Lagrangian is not known or cannot be constructed.

As shown in Ref.~\cite{Slaughter:2011xs} and references therein, infinite-momentum frame broken asymptotic symmetry is characterized by the existence of physical on-mass-shell hadron annihilation
operators $a_{\alpha }(\vec{k},\lambda )$ (momentum
$\vec{k}$ ($|\vec{k}|\rightarrow \infty $), helicity $\lambda $,
and $SU_{F}(N)$ flavor index $\alpha $) and their creation operator counterparts which produce physical states when acting on the vacuum.  Indeed, the \emph{physical} on-mass-shell hadron annihilation
operator $a_{\alpha }(\vec{k},\lambda )$
 is related linearly under flavor transformations to the {\em {representation}} annihilation operator
$a_{j}(\vec{k},\lambda )$. Thus, in the infinite-momentum frame, physical states denoted by $\vert \alpha,\vec k, \lambda\rangle$ ( which do not belong to irreducible representations) are linear combinations of representation states denoted by $\vert j, \vec k,\lambda\rangle$ (which do belong to irreducible representations) plus nonlinear corrective terms that are best calculated in a frame where mass differences are deemphasized such as in the infinite-momentum frame.  Mathematically \cite{Oneda:1985wf,Oneda:1991wz,Slaughter:2011xs,Slaughter:2008zd,Oneda:1979mp,SlaughterOneda:1977PRL,YasueOnedaSlaughter:1983PRL,SlaughterOneda:1987PRL}, this is expressed by: $\vert \alpha,\vec k, \lambda\rangle=\sum_j C_{\alpha j}\vert j, \vec k,
\lambda\rangle$, $\vert \vec k \vert \rightarrow\infty$,
where the orthogonal matrix $C_{\alpha j}$ depends on \emph{physical} $SU_F(N)$ mixing
parameters, is defined \emph{only} in the $\infty$-momentum frame, and can be constrained directly by ETCRs.

  It cannot be overemphasized that the particular Lorentz frame that one utilizes when analyzing current-algebraic sum rules does not matter when flavor symmetry
is exact and is strictly a matter of taste and calculational convenience, whereas when one uses current-algebraic sum rules in broken symmetry, the choice of frame is \emph{paramount} since one wishes to emphasize the calculation of leading order contributions while simultaneously simplifying the calculation of symmetry breaking corrections \cite{Oneda:1985wf,Oneda:1991wz,Slaughter:2011xs,Slaughter:2008zd,Oneda:1989jr,Oneda:1989ik,Slaughter:1988hx,
Oneda:1979mp,Perl:1974,Carruthers:1971,alfaro63}.

While we will only discuss the $J^{PC}=0^{-+}$ $35$-plet representation in this paper, nevertheless it is instructive to outline our normalization conventions including fermionic representation states. In Table~5, we give all non-zero anti-commutation relations, where the singlet $U(1)$ matrix $V_{0}$ is explicitly present. We have: particle four-momenta $p=(p^{0},\vec{p})$ and $p'=(p'^{\, 0},\vec{p'})$, with ${\left[a^{(r)}(p),a^{\dag(s)}(p')\right]}_{+}=
{\left[b^{(r)}(p),b^{\dag(s)}(p')\right]}_{+}$$=$
$N_{a} \, \delta_{rs}\delta^{3}(\vec{p}-\vec{p'})$,
$u^{\dag (r)}(p)\,u^{(s)}(p)$ $=$ $N_{D}\, \delta_{rs}$,
\newline $\psi(x)=\sum_{r}\int \mathrm{d^{3}p}\, N_{\psi}\left[a^{(r)}(p)u^{(r)}(p)\,e^{-ip\cdot x}+b^{\dag(r)}(p)v^{(r)}(p)\;e^{+ip\cdot x}\right]$,
\newline $a^{(r)}(p)=((2\pi)^{3}N_{\psi}N_{D})^{-1}\int \mathrm{d^{3}x}\,e^{+ip\cdot x}\,\bar{u}^{r}(p)\gamma_{0}\,\psi(x)$,
$b^{(r)}(p)=((2\pi)^{3}N_{\psi}N_{D})^{-1}\int \mathrm{d^{3}x}\,e^{+ip\cdot x}\,\bar{\psi}(x)\gamma_{0}\,v^{(r)}(p)$,
\newline  $\brkt{s',p',\lambda'}{s,p,\lambda}$ $=$ $\delta_{ss'}\,\delta_{\lambda \lambda'}\,{(}2\pi {)^{3}}\,2p^{\;0}\,\delta^{3}(\vec{p^{\prime}}\,-\vec{p})$, where
 $a^{(r)}(p)$ and $b^{(r)}(p)$ are creation operators, $u^{(s)}(p)$, $v^{(r)}(p)$ \\ are Dirac spinors, $\psi(x)$ is a spin $1/2$ Dirac field operator ($s$, $\lambda$ denote particle spin and helicity respectively),\\ $\ket{\psi}=\sum_{s,\lambda}\int N_{a}^{-1}\mathrm{d^{3}p}\,\ket{s,p,\lambda}\brkt{s,p,\lambda}{\psi}$, $ (2\pi)^{6} N_{\psi}^{2}N_{D}^{2} \braket{N}{\Omega^{\mu\ldots}}{N}$ is covariant (transforms like $\Omega^{\mu\ldots}$),\\ and $(2\pi)^{3}N_{\psi}^{2}N_{D}N_{a}=1$.

\begin{center}
 {\large\textbf{III. THE PHYSICAL ELECTROMAGNETIC CURRENT}}
\end{center}

We now discuss the vector charges and two-particle basis states after imposing a unitary homogeneous pure Lorentz transformation $-\hat{z}$ boost such that the $3$-momentum $\vec{k}$ of all states has
 $|\vec{k}|\rightarrow \infty $,
 and all creation operators produce \emph{physical} states when acting on the physical vacuum, See~ Refs. (\cite{Slaughter:2011xs,Slaughter:2008zd}) for more details.

 First, we note that the charges operating on meson states [See Equations~(\ref{eq:roots},\ref{eqbasic8}, and \ref{eqbasic9})] transform according to :
\begin{equation}\label{eq:stategen}
V_{i}^{j}\ket{q_{k}\bar{q}_{l}}=\delta_{k}^{j}\ket{q_{i}\bar{q}_{l}}-\delta_{i}^{l}\ket{q_{k}\bar{q}_{j}}.
\end{equation}
Kets (four-momentum $p$, quantum numbers given by their quark-antiquark content), and bras (four-momentum $p'$) are given by:
\begin{equation}\label{eq:stategen1}
  \ket{{D}^{+}, p}\,(J^{PC}=0^{-+})=\ket{c\bar{d}}=\ket{u_{4}\bar{u}_{2}},\quad  \textrm{and} \, \bra{{D^{\ast}}^{+}, p'}\,(J^{PC}=1^{--})\equiv\bra{^{*}c\bar{d}}=\bra{^{*}u_{4}\bar{u}_{2}}\quad \, \emph{etc.} .
\end{equation}
where the space-like four-momentum transfer $q^{2}$ is given by:
\begin{eqnarray}
  &&q^{2}  = (p'-p)^{2}={m^{*}}^{2}+{m}^{2}-2 p' \cdot p\,,\label{eq:momxfer1}\\
  &&p' = (E',\vec{s} )=(\sqrt{{m^{*}}^{2}+s_x^2+s_z^2},s_x,0,s_z )\,,\label{eq:momxfer2}\\
  &&p = (E,\vec{t} )=(\sqrt{m^2+t_z^2},0,0,t_z )\,,\\
  && \textsl{Set} \,s_{z}=r t_{z}\quad \textrm{and}  \quad\mbox{$0<r=\textsl{constant} $\,\,} \,, \label{eq:momxfer3}\\
  &&p' \cdot p=\sqrt{{m^{*}}^{2}+s_x^2+r^2 t_z^2}\,\sqrt{m^2+t_z^2}-rt_z^2 \,,\label{eq:momxfer4}\\
  &&p' \cdot p=rt_z^2\sqrt{1+\frac{{m^{*}}^{2}+s_x^2}{r^2 t_z^2}}\sqrt{1+\frac{m^2+t_z^2}{t_z^2}}-rt_z^2 \,,\label{eq:momxfer5}\\
  &&\stackrel{t_z\rightarrow \infty}{\overbrace{p' \cdot p}}\, = \frac{1}{2}\left(\frac{{m^{*}}^{2}+s_x^2+r^2 m^2}{r}\right)\,,\label{eq:momxfer6}\\
  &&q^{2}  ={m^{*}}^{2}+ m^2-\left(\frac{{m^{*}}^{2}+s_x^2+r^2 m^2}{r}\right)\,,\label{eq:momxfer7}\\
  &&q^{2}  =\frac{-(1-r)}{r}\,{m^{*}}^{2}\, [1- (\frac{m^{2}}{{m^{*}}^{2}}) \,r ] - \frac{s_x^2}{r}\,.\label{eq:momxfer8}
 \end{eqnarray}

 The physical vector charge $V_{K^{0}}$ is $V_{K^{0}}=V_{6}+iV_{7}$, the physical vector charge $V_{\pi^{\pm}}=V_{1}\pm iV_{2}$, \textit{etc.}. The $\lambda_{a}$,
$a=1,2,\cdots,35$ satisfy the Lie algebra
$[(\lambda_{a}/2),(\lambda_{b}/2)]=i\sum_{c}f_{abc}(\lambda_{c}/2),$ where the $f_{abc}$ are
structure constants of the flavor group ${SU_{F}(6)}$ and ${{V_{a}}^{\mu}}(x)=\bar{q}^{i}(x){(\lambda_{a}/2)}_{ij}
\gamma^{\mu}q^{j}(x)$.


 The \emph{physical electromagnetic current} $j_{em}^{\mu}(0)$ ($u$, $d$%
, $s$, $c$, $b$, $t$ quark system) is:

\begin{align}\label{eq:electromagnetic}
  j_{em}^{\mu}(0) =& V_{3}^{\mu}(0)+(1/3)^{1/2}V_{8}^{\mu}(0)-(2/3)^{1/2}V_{15}^{\mu}(0) \\
  +&(2/5)^{1/2}V_{24}^{\mu}(0)- (3/5)^{1/2}V_{35}^{\mu}(0)\nonumber\\
  +&(1/3)^{1/2}\times\mbox{${{V_{0}}^{\mu}}(0)$}\nonumber\\
  =&j^{\mu}_{V}(0)+j^{\mu}_{S}(0)\, ,
\end{align}

where $j^{\mu}_{V}(0)\equiv j^{\mu}_{em\;3}(0)=V_{3}^{\mu}(0)$ is the iso-vector part of the electromagnetic current, $j^{\mu}_{S}(0)\equiv$ the iso-scalar part of the electromagnetic current.   The flavor $U(1)$ singlet current ${{V_{0}}^{\mu}}(x)=\bar{q}^{i}(x){(\lambda_{0}/2)}_{ij}\gamma^{\mu}q^{j}(x)$ where $\lambda_{0}\equiv\sqrt{1/3} \;I$, $I$ is the identity, so that $\mathrm{Tr}(\lambda_{a}\lambda_{b})=2\delta_{ab}$ holds for all $\lambda_{a'}$ ($a'=0,1,2,\ldots,35)$, and $j^{\mu }_{0}=V_0^{\mu }/\sqrt{3}$\,. The $U(1)$ singlet charge $V_{0}$ commutes with all of the $V_{a}$.\\

 From the commutation relations in Table 3, \emph{we obtain the
  fascinating equation [it contains only ETCRs \emph{and} explicitly the electromagnetic current singlet $j^{\mu }_{0}$ ]--in broken symmetry} [$j^{\mu }\equiv j^{\mu}_{em}(0)$ (momentum
$\vec{k}$ with $|\vec{k}|\rightarrow \infty $):

\begin{equation}\label{eqMain}
  [[j^{\mu },V_{\pi ^+}],V_{\pi ^-}]+[[j^{\mu },V_{D_{s}^{-}}],V_{D_{s}^{+}}]+[[j^{\mu },V_{T_{b}^{-}}],V_{T_{b}^{+}}]=2 j^{\mu }-2 (V_0^{\mu }/\sqrt{3})=2 j^{\mu }- 2 j^{\mu }_{0}
\end{equation}
\\

The angles between the simple roots appearing in Eq.~(\ref{eqMain}) are: $\theta_{\vec{\rho}\,(\pi^{+})  ,\vec{\rho}\,(D_{s}^{-})} = \theta_{\vec{\rho}\,(D_{s}^{-})  ,\vec{\rho}\,(T_{b}^{-})}=90\,^{\circ}$.  These angles correspond to the $(u,d)$, $(c,s)$, and $(t,b)$ doublet ``sectors".

For vector ($V$) $\rightarrow$ pseudoscalar ($P$) radiative decays, we have (in the infinite-momentum frame [$t_{z}\rightarrow \infty$, see Eq.~(\ref{eq:momxfer8})],\\ $\lambda=1$= vector meson polarization index, $\mu=0$, $r=1$, and $s_{x}^{2}=0$  $\Rightarrow q^{2}=0$ in the following matrix element):
\begin{subequations}\label{eq:raddecay1}
\begin{equation}\label{eq:radcoupling}
  \braket{{V}}{j^{\mu}}{{P}}=\brak{V}\epsilon^{\mu\nu\rho\sigma}\epsilon_{\nu}(\vec{p},\lambda)p'_{\rho}\, p_{\sigma}
\end{equation}
\begin{equation}
 \Gamma(V({p',\lambda=1)}\rightarrow P({p)}\,+ \gamma (q^{2}=0)) =\frac{{\brak{V}}^{2}}{96\pi}  \left({{\frac{ m^{2}_{V}-m^{2}_{P} }{m_{V}}}}\right)^{3}
\end{equation}
\begin{equation}\label{eq:raddecay}
  |\brak{V}|=\left[{96\pi} \left({{\frac{ m_{V} }{   m^{2}_{V}-m^{2}_{P}   }}}\right)^{3}\Gamma(V \rightarrow P + \gamma)\right]^{\frac{1}{2}}
\end{equation}
\end{subequations}

So for instance, if we evaluate Eq.~(\ref{eqMain}) between the states $\bra{{D^{\ast}}^{+}}$ and $\ket{{D}^{+}}$ using
Equations~(\ref{eqbasic8}), (\ref{eq:stategen}), and (\ref{eq:stategen1}) we obtain:

\begin{equation}\label{eq:singlet}
\brak{{D^{\ast}}^{0}}+\brak{   {{{\overline{K}}}^{*}}^{0}    }=2 \brak{{{D^{\ast}}^{+}}}_{0}\, ,
\end{equation}

where $\brak{{D^{\ast}}^{0}}\equiv \braket{{D^{\ast}}^{0}}{j^{\mu}}{{D}^{0}}$,
$\brak{   {{{\overline{K}}}^{\ast}}^{0}}\equiv \braket{{{\overline{K}}^{\ast}}^{0}}{j^{\mu}}{{K}^{0}}$,
${\brak{{\rho}^{+}}_{0}\equiv \braket{{\rho}^{+}}{\frac{1}{\sqrt{3}}V^{\mu}_{0}}{{\pi}^{+}}}$,
and ${\brak{{D^{\ast}}^{0}}_{0}\equiv \braket{{D^{\ast}}^{0}}{\frac{1}{\sqrt{3}}V^{\mu}_{0}}{{D}^{0}}}$,  \textit{etc.}.  \\
\vspace{0.5in}

We define:
\begin{equation}\label{eq:stack}
\stackMath
\savestack{\pdqvec}{  \Longstack{ {\rho^{0}} {\omega} {\phi} {\psi} {\Upsilon} {t\overline{t}}}}%
\savestack{\pdqpsc}{  \Longstack{  {\eta} {\eta'} {\eta_{c}} {\eta_{b}} {\eta_{t}}}}%
X_{1^{--}}\equiv\begin{pmatrix}\pdqvec\end{pmatrix} \,\,  \, \,
\end{equation}
\vspace{0.5in}
\begin{center}
 \textrm{and}
\end{center}

\vspace{0.5in}
\begin{equation}\label{eq:stack1}
\stackMath
\savestack{\pdqpsc}{  \Longstack{ {\pi^{0}} {\eta} {\eta'} {\eta_{c}} {\eta_{b}} {\eta_{t}}}}%
X_{0^{-+}}\equiv\begin{pmatrix}\pdqpsc\end{pmatrix} \,.
\end{equation}
\\
\vspace{0.5in}
Evaluating Eq.~(\ref{eqMain}) between the $16$ bra-ket state pairs with bras:

\begin{gather}
\bra{{\rho}^{0}}\, ,\bra{{\rho}^{+}}\, ,\bra{{K^{\ast}}^{+}}\, ,\bra{{\overline{D}^{\ast}}^{0}}\, ,\nonumber \\
\bra{{B^{\ast}}^{+}}\,  ,\bra{{\overline{T}^{\ast}}^{0}},\bra{{K^{\ast}}^{0}}\, ,\bra{{D^{\ast}}^{-}}\, ,\nonumber \\
\bra{{B^{\ast}}^{0}}\, ,\bra{{T^{\ast}}^{-}}\, ,\bra{{{{D}_{s}^{\ast}}}^{-}}\, ,\bra{{{{B}_{s}^{\ast}}}^{0}}\, ,\nonumber \\
\bra{{{{T}_{s}^{\ast}}}^{-}}\, ,\bra{{{{B}_{c}^{\ast}}}^{+}}\, ,\bra{{{{\overline{T}}_{c}^{\ast}}}^{0}}\, , \textrm{and} \bra{{{{\overline{T}}_{b}^{\ast}}}^{-}}\, ,\nonumber
\end{gather}
\vspace{0.5in}
then we find that:
\begin{align}
  \brak{{\rho}^{+}} =\brak{{\rho}^{-}} =\brak{{\rho^{0}}}_{0}\,
  ,\label{eq:derive161test}\\
  \brak{{\rho}^{0}}= \brak{{{\rho}^{+}}}_{0}\,
  ,\label{eq:derive162test}\\
  \brak{{K^{\ast}}^{0}}+\brak{   {{{\overline{D}}}^{*}}^{0}    }=2 \brak{{{K^{\ast}}^{+}}}_{0}\, ,\label{eq:derive163test}\\
  \brak{{K^{\ast}}^{+}}+\brak{   {{{D}}^{*}}^{-}    }=2 \brak{  {{{\overline{D}}}^{*}}^{0}   }_{0}\, ,\label{eq:derive164test}\\
   \brak{{B^{\ast}}^{0}}+   \brak{   {{{\overline{T}}}^{*}}^{0}    }=2 \brak{{{B^{\ast}}^{+}}}_{0}\, ,\label{eq:derive165test}\\
  \brak{{B^{\ast}}^{+}}+   \brak{   {{{T}}^{*}}^{-}    }=2 \brak{  {{{\overline{T}}}^{*}}^{0}   }_{0}\, ,\label{eq:derive166test}\\
  \brak{{K^{\ast}}^{+}}+ \brak{   {{{D}}^{*}}^{-}    }=2 \brak{{{K^{\ast}}^{0}}}_{0}\,
  ,\label{eq:derive167test}\\
  \brak{{K^{\ast}}^{0}}+ \brak{   {{{{{\overline{D}}}}}^{*}}^{0}    }=2 \brak{{{D^{\ast}}^{-}}}_{0}\, ,\label{eq:derive168test}\\
   \brak{{B^{\ast}}^{+}}+ \brak{   {{{T}}^{*}}^{-}    }=2 \brak{{{B^{\ast}}^{0}}}_{0}\,
   ,\label{eq:derive169test}\\
   \brak{{B^{\ast}}^{0}}+ \brak{   {{{\overline{T}}}^{*}}^{0}       }=2 \brak{{{T^{\ast}}^{-}}}_{0}\, ,\label{eq:derive1610test}\\
   \brak{{{{D}_{s}^{\ast}}}^{-}}\braket{{{{D}_{s}}}^{-}}{V_{D_{s}^{-}}}{ \bigg[  X_{0^{-+}} \,  }\braket{ X_{0^{-+}}  \bigg] \,}{V_{{{D}_{s}}}^{+}}{ \,{{{D}_{s}}}^{-} }-
  \nonumber \\
  \braket{ \, {{{D}_{s}^{\ast}}}^{-}   \,}{V_{{D}_{s}^{-}}}{\bigg[\,  \,X_{1^{--}}   \, }
  \braket{ \,  X_{1^{--}}  \bigg] \,}{{j}^{\mu}}{\bigg[ \,X_{0^{-+}} }\braket{ \, X_{0^{-+}}  \bigg] \,}{V_{{{D}_{s}}}^{+}}{ \,{{{D}_{s}}}^{-} }
   \nonumber \\
  =2 \brak{{{{D}_{s}^{\ast}}}^{-}}  -2 \brak{{{{D}_{s}^{\ast}}}^{-}}_{0}  \,
  ,\label{eq:derive1611test}\\
   \brak{{{{B}_{c}^{\ast}}}^{+}}+ \brak{{{{T}_{s}^{\ast}}}^{-}}=2 \brak{{{{B}_{s}^{\ast}}}^{0}}_{0}\, ,\label{eq:derive1612test}\\
   \brak{{{{B}_{s}^{\ast}}}^{0}}+ \brak{{{{\overline{T}}_{c}^{\ast}}}^{0}}=2 \brak{{{{T}_{s}^{\ast}}}^{-}}_{0}\, ,\label{eq:derive1613test}\\
   \brak{{{{B}_{s}^{\ast}}}^{0}}+ \brak{{{{\overline{T}}_{c}^{\ast}}}^{0}}=2 \brak{{{{B}_{c}^{\ast}}}^{+}}_{0}\, ,\label{eq:derive1614test}\\
   \brak{{{{B}_{c}^{\ast}}}^{+}}+ \brak{{{{\overline{T}}_{s}^{\ast}}}^{-}}=2 \brak{{{{\overline{T}}_{c}^{\ast}}}^{0}}_{0}\,
   ,\label{eq:derive1615test}\\
   \brak{{{{T}_{b}^{\ast}}}^{-}}\braket{{{{T}_{b}}}^{-}}{V_{{{{T}_{b}}}^{-}}}{ \bigg[ \,X_{0^{-+}}\,  }\braket{ \,X_{0^{-+}} \bigg] \,}{V_{T_{b}^{+}}}{ \,{{{T}_{b}}}^{-}\, }-
  \nonumber \\
   \braket{ \, {{{T}_{b}^{\ast}}}^{-}   \,}{V_{{T}_{b}^{-}}}{\bigg[\,  \, X_{1^{--}}   \, }
  \braket{ \, X_{1^{--}} \bigg] \,}{{j}^{\mu}}{\bigg[ \,X_{0^{-+}} }\braket{ \,X_{0^{-+}} \bigg] \,}{V_{T_{b}^{+}}}{ \,{{{T}_{b}}}^{-} }
   \nonumber \\
  =2 \brak{{{{T}_{b}^{\ast}}}^{-}}  - 2\brak{{{{T}_{b}^{\ast}}}^{-}}_{0}  \,
  .\label{eq:derive1616test}
\end{align}

Equations~(\ref{eq:derive161test}) to (\ref{eq:derive1616test}) \emph{explicitly demonstrate in broken symmetry} the importance of the electromagnetic current singlet $U(1)$ matrix $V_{0}$ contribution to radiative decays}.  Indeed, one finds that $\Gamma ( \rho^{\pm}\rightarrow \pi^{\pm}\,\gamma)$ and  $\Gamma ( \rho^{0}\rightarrow \pi^{0}\,\gamma)$) are \emph{entirely due to the electromagnetic current singlet contribution}.  At present, insufficient data is available for most of the decay matrix elements in Equations~(\ref{eq:derive161test}) to (\ref{eq:derive1616test}).  Even where there is data, the signs of the matrix elements are not yet experimentally available, although there exist theoretical models which predict matrix element signs \cite{ODonnell:1981hgt}.  From Equation~(\ref{eq:derive163test}) and Equation~(\ref{eq:derive168test}), we find that $\brak{{{D^{\ast}}^{-}}}_{0}=\brak{{{K^{\ast}}^{+}}}_{0}$. Similarly, we find from Equation~(\ref{eq:derive164test}) and Equation~(\ref{eq:derive167test}) we find that $\brak{  {{{\overline{D}}}^{*}}^{0}   }_{0}=\brak{{{K^{\ast}}^{0}}}_{0}$. Very little is known about the behavior of the singlet generator in broken symmetry, other than that given by Equations~(\ref{eq:derive161test}) to (\ref{eq:derive1616test})---to partially remedy that situation let the right-hand sides of Equations~(\ref{eq:derive163test}) and (\ref{eq:derive167test}) be proportional---\textit{i.e.} $\brak{{{K^{\ast}}^{0}}}_{0}=\beta\brak{{{K^{\ast}}^{+}}}_{0}$.  We then obtain\cite{Tanabashi:2018xqp,Slaughter:charge conjugate states}:
\begin{equation}\label{eq:D0KPlus)}
 \frac{\brak{ {{{D}}^{*}}^{0}  }}{ \brak{ {K^{\ast}}^{+}}} = \frac{1}{\beta}*[1 + \frac{\brak{ {{{D}}^{*}}^{-}  }}{\brak{ {K^{\ast}}^{+}}}-\beta*\frac{\brak{ {{{K}}^{*}}^{0}  }}{\brak{ {K^{\ast}}^{+}}}]\qquad \textrm{where} \quad\beta\neq 0.
\end{equation}

From data in Ref.~( \cite{Tanabashi:2018xqp}), we get ($\pm$ signs are not correlated and $\beta$ is assumed to be $1$) [an assumption suggested only by  Equations~(\ref{eq:derive161test}) and (\ref{eq:derive162test}) $\rho$ triplet charged and neutral singlet results and perhaps holding for doublets as well]:
\begin{equation}\label{eq:D0KPlusNum}
  \frac{\brak{ {{{D}}^{*}}^{0}  }}{ \brak{ {K^{\ast}}^{+}}} = 1+(\pm (0.56\pm0.08))-(\pm (1.52\pm 0.10))=
   \left\{
\begin{array}{rll}
+0.05 \pm 0.13 & \text{ for }\, ++\\
+3.06 \pm 0.13 & \text{ for }\, +-\\
-1.06 \pm 0.13 & \text{ for }\, -+ \\
+1.95 \pm 0.13 & \text{ for }\, --
\end{array} \right.
\end{equation}
This implies that (statistical propagation of errors---quadrature calculated):

\begin{equation}
 \Gamma({{{D}}^{*}}^{0}\rightarrow D^{0}\,\gamma)_{\beta =1}= \left\{
\begin{array}{rll}
0.13^{\, +0.65}_{\,-0.13} & \text{keV for }\, ++\\
468. \pm 61. & \text{keV for }\, +-\\
56. \pm 15. & \text{keV for }\, -+ \\
189. \pm 31. & \text{keV for }\, --
\end{array} \right.
\end{equation}

\begin{figure}[h]
\centering
\includegraphics{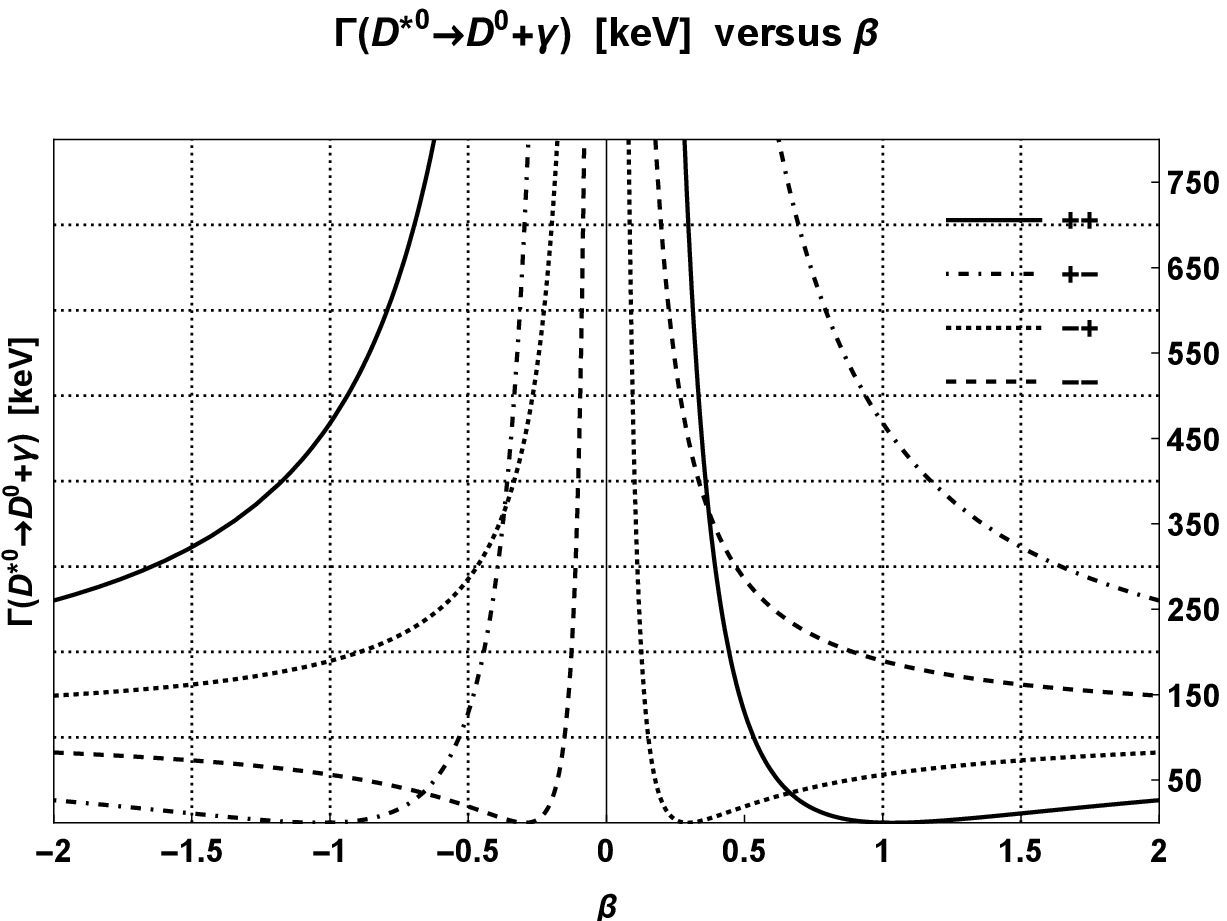}
\caption{{\footnotesize $\mathbf{\Gamma({{{D}}^{*}}^{0}\rightarrow D^{0}\,+ \,\boldsymbol{\gamma})}$ \textbf{versus}  $\boldsymbol{\beta}$
}} \label{fig1}
\end{figure}

In Figure~\ref{fig1}, we graphicly show $\Gamma({{{D}}^{*}}^{0}\rightarrow D^{0}\,+ \,\gamma)$ versus  $\beta$ as $\beta$ is allowed to vary from $-2$ to $2$.
\vspace{1cm}

\vspace{1cm}

From Figure 1. above, we see that $\Gamma({{{D}}^{*}}^{0}\rightarrow D^{0}\,\gamma)$ is dependent upon the signs of $\frac{\brak{ {{{D}}^{*}}^{-}  }}{\brak{ {K^{\ast}}^{+}}}$ and $\frac{\brak{ {{{K}}^{*}}^{0}  }}{\brak{ {K^{\ast}}^{+}}}$.  However, in unbroken $SU(3)$ $\frac{\brak{ {{{K}}^{*}}^{0}  }}{\brak{ {K^{\ast}}^{+}}}$ is negative and $=-2$\cite{ODonnell:1981hgt,Oneda:1991wz,Slaughter:1988hx}. From experiment we therefore choose $\frac{\brak{ {{{K}}^{*}}^{0}  }}{\brak{ {K^{\ast}}^{+}}}=-1.51\pm 0.10$, so we expect the $+-$ or $--$ curves in Fig.1 above to be more physically predictive.  From Table 3, we have at our disposal:
\begin{subequations}\label{eq:twocr}
\begin{equation}\label{twocr1}
  [V_{K^{0}},j^{\mu}]=0 \quad \textrm{and}
\end{equation}
\begin{equation}\label{twocr2}
  [V_{D^{0}},j^{\mu}]=0.
\end{equation}
\end{subequations}
Evaluating (we neglect intermultiplet mixing with the $K^{*}(1410)$ ) Equation~(\ref{twocr1}) between the physical asymptotic states $\bra{{K^{+}}}$ and $\ket{{\rho^{+}}}$ and Equation~(\ref{twocr2}) between the physical asymptotic states $\bra{{D^{+}}}$ and $\ket{{\rho^{+}}}$, one obtains
\begin{equation}\label{onesign}
  \brak{ {D^{\ast}}^{+} }=\brak{ {K^{\ast}}^{+} }=\brak{ {\rho}^{+} }.
\end{equation}
Thus, we expect (conjugate states used) that $\frac{\brak{ {{{D}}^{*}}^{-}  }}{\brak{ {K^{\ast}}^{+}}}$ is positive and \emph{the $+-$ curve in Figure 1. to be that which is operative}. An experimental determination of $\Gamma({{{D}}^{*}}^{0}\rightarrow D^{0}\,\gamma)$ would then provide a value for $\beta$ which may be useful in further research--especially where higher rank special unitary groups and Lie algebras play a role.
Unfortunately, current experimental data from \cite{Tanabashi:2018xqp} yields only that $\Gamma({{{D}}^{*}}^{0}\rightarrow D^{0}\,\gamma)\leq 741.3 \, \text{keV, \, CL=90\%}$.

At present, quantum field theories (including SUSY theories) have not been successful in replacing the standard model (QCD...which does not include gravity) with a grand unified theory without hierarchical or other problems. Generally speaking, most theories are perturbative and renormalizable with local gauge fields strongly related to Lie Algebras and utilize spontaneous symmetry breaking.  In Lie algebraic representations of interest, anomalies (see Ref.~{\cite{georgi:1999lie} and Ref.~(\cite{weinberg:vol2} and Ref.~{\cite{Alvarez-Gaume:2013CERN}})--especially chapter 22} for instance) and Refs.~(\cite{AlvarezGaume:1985ex},\cite{Bilal:2008qx}) must vanish for physical representations---a severe constrain on those theories. On the other hand, Equations~(\ref{eq:derive161test}) to (\ref{eq:derive1616test}) are \emph{non-perturbative}.

\begin{center}
 {\large\textbf{IV. SUMMARY AND CONCLUSIONS}}
\end{center}

We presented research on radiative decays of vector ($J^{PC}=1^{--}$) to pseudoscalar ($J^{PC}=0^{-+}$) particles ($u$, $d$, $s$, $c$, $b$, $t$ quark system) using broken symmetry techniques in the infinite momentum frame and equal time commutation relations. The research utilized the $SU(6)$ Lie algebra characterization of  flavor $SU_{F}(6)$ \emph{representations} and the \emph{physical electromagnetic current} $j_{em}^{\mu}(0)$ including its singlet $U(1)$ term and focused on the $35$-plet. The research was conducted without ascribing any specific form to meson quark structure or intra-quark interactions by using \emph{equal-time commutation relations }(ETCRs) which involve at most one current density, thus, \emph{avoiding problems associated with Schwinger terms}. We found that \emph{the electromagnetic current singlet plays an intrinsic role in understanding the physics of radiative decays} where (bilinear) commutators of the $SU(6)$ Lie algebra generators are Lie products acting over the real number field. Indeed, in broken symmetry and the infinite momentum frame, we developed a new and fascinating equation involving the electromagnetic current (including its singlet--proportional to the $SU(6)$ singlet), three $SU(6)$ simple roots, and double commutators using ETCRs.

For notational conciseness and self-containment and use by other researchers, $SU(6)$ Lie algebra simple roots, positive roots, weights, fundamental weights, non-zero commutators, and non-zero anti-commutators were also determined which allow construction of all $SU(6)$ representations. Surprisingly\textemdash after symmetry breaking\textemdash we discovered that charged and neutral $\rho$ meson radiative decays into $\pi \, \gamma$ were due entirely to the singlet term in $j_{em}^{\mu}(0)$. Although there is insufficient experimental data on the radiative decay $\Gamma({{{D}}^{*}}^{0}\rightarrow D^{0}\,\gamma)$ available at this time, we derived equations involving \emph{physical} matrix elements of the $SU(6)$ singlet generator which allowed parametrization of possible predicted values of $\Gamma({{{D}}^{*}}^{0}\rightarrow D^{0}\,\gamma)$ versus $\beta$.

\newpage
{\footnotesize
\noindent
\begin{table}[!ht]
  {\bf{Table 1:}   }
\begin{center}
\section*{``Generalized" Gell-Mann and Flavor $\mathbf{U(1)}$ Singlet Matrices   }
\end{center}
\doublerule

\begin{minipage}{3cm}
\scriptsize
$$\lambda_{1}=
\left(
\begin{array}{cccccc}
 0 & 1 & 0 & 0 & 0 & 0 \\
 1 & 0 & 0 & 0 & 0 & 0 \\
 0 & 0 & 0 & 0 & 0 & 0 \\
 0 & 0 & 0 & 0 & 0 & 0 \\
 0 & 0 & 0 & 0 & 0 & 0 \\
 0 & 0 & 0 & 0 & 0 & 0
\end{array}
\right),
$$
\end{minipage} \qquad \qquad \qquad
\begin{minipage}{3cm}
\scriptsize
$$\lambda_{2}=
\left(
\begin{array}{cccccc}
 0 & -i & 0 & 0 & 0 & 0 \\
 i & 0 & 0 & 0 & 0 & 0 \\
 0 & 0 & 0 & 0 & 0 & 0 \\
 0 & 0 & 0 & 0 & 0 & 0 \\
 0 & 0 & 0 & 0 & 0 & 0 \\
 0 & 0 & 0 & 0 & 0 & 0
\end{array}
\right),
$$
\end{minipage} \qquad \qquad \qquad
\begin{minipage}{3cm}
\scriptsize
$$\lambda_{3}=
\left(
\begin{array}{cccccc}
 1 & 0 & 0 & 0 & 0 & 0 \\
 0 & -1 & 0 & 0 & 0 & 0 \\
 0 & 0 & 0 & 0 & 0 & 0 \\
 0 & 0 & 0 & 0 & 0 & 0 \\
 0 & 0 & 0 & 0 & 0 & 0 \\
 0 & 0 & 0 & 0 & 0 & 0
\end{array}
\right),
$$
\end{minipage}, \qquad \qquad \qquad
\\
\begin{minipage}{3cm}
\scriptsize
$$\lambda_{4}=
\left(
\begin{array}{cccccc}
 0 & 0 & 1 & 0 & 0 & 0 \\
 0 & 0 & 0 & 0 & 0 & 0 \\
 1 & 0 & 0 & 0 & 0 & 0 \\
 0 & 0 & 0 & 0 & 0 & 0 \\
 0 & 0 & 0 & 0 & 0 & 0 \\
 0 & 0 & 0 & 0 & 0 & 0
\end{array}
\right),
$$
\end{minipage} \qquad \qquad \qquad
\begin{minipage}{3cm}
\scriptsize
$$\lambda_{5}=
\left(
\begin{array}{cccccc}
 0 & 0 & -i & 0 & 0 & 0 \\
 0 & 0 & 0 & 0 & 0 & 0 \\
 i & 0 & 0 & 0 & 0 & 0 \\
 0 & 0 & 0 & 0 & 0 & 0 \\
 0 & 0 & 0 & 0 & 0 & 0 \\
 0 & 0 & 0 & 0 & 0 & 0
\end{array}
\right),
$$
\end{minipage}\qquad \qquad \qquad
\begin{minipage}{3cm}
\scriptsize
$$\lambda_{6}=
\left(
\begin{array}{cccccc}
 0 & 0 & 0 & 0 & 0 & 0 \\
 0 & 0 & 1 & 0 & 0 & 0 \\
 0 & 1 & 0 & 0 & 0 & 0 \\
 0 & 0 & 0 & 0 & 0 & 0 \\
 0 & 0 & 0 & 0 & 0 & 0 \\
 0 & 0 & 0 & 0 & 0 & 0
\end{array}
\right),
$$
\end{minipage}\qquad \qquad \qquad
\\
\begin{minipage}{3cm}
\scriptsize
$$\lambda_{7}=
\left(
\begin{array}{cccccc}
 0 & 0 & 0 & 0 & 0 & 0 \\
 0 & 0 & -i & 0 & 0 & 0 \\
 0 & i & 0 & 0 & 0 & 0 \\
 0 & 0 & 0 & 0 & 0 & 0 \\
 0 & 0 & 0 & 0 & 0 & 0 \\
 0 & 0 & 0 & 0 & 0 & 0
\end{array}
\right),
$$
\end{minipage}\qquad \qquad \qquad
\begin{minipage}{3cm}
\scriptsize
$$\lambda_{8}=
\left(
\begin{array}{cccccc}
 \frac{1}{\sqrt{3}} & 0 & 0 & 0 & 0 & 0 \\
 0 & \frac{1}{\sqrt{3}} & 0 & 0 & 0 & 0 \\
 0 & 0 & -\frac{2}{\sqrt{3}} & 0 & 0 & 0 \\
 0 & 0 & 0 & 0 & 0 & 0 \\
 0 & 0 & 0 & 0 & 0 & 0 \\
 0 & 0 & 0 & 0 & 0 & 0
\end{array}
\right),
$$
\end{minipage}\qquad \qquad \qquad \qquad
\begin{minipage}{3cm}
\scriptsize
$$\lambda_{9}=
\left(
\begin{array}{cccccc}
 0 & 0 & 0 & 1 & 0 & 0 \\
 0 & 0 & 0 & 0 & 0 & 0 \\
 0 & 0 & 0 & 0 & 0 & 0 \\
 1 & 0 & 0 & 0 & 0 & 0 \\
 0 & 0 & 0 & 0 & 0 & 0 \\
 0 & 0 & 0 & 0 & 0 & 0
\end{array}
\right),
$$
\end{minipage}\qquad \qquad \qquad \qquad
\\
\begin{minipage}{3cm}
\scriptsize
$$\lambda_{10}=
\left(
\begin{array}{cccccc}
 0 & 0 & 0 & -i & 0 & 0 \\
 0 & 0 & 0 & 0 & 0 & 0 \\
 0 & 0 & 0 & 0 & 0 & 0 \\
 i & 0 & 0 & 0 & 0 & 0 \\
 0 & 0 & 0 & 0 & 0 & 0 \\
 0 & 0 & 0 & 0 & 0 & 0
\end{array}
\right),
$$
\end{minipage}\qquad \qquad \qquad \qquad
\begin{minipage}{3cm}
\scriptsize
$$\lambda_{11}=
\left(
\begin{array}{cccccc}
 0 & 0 & 0 & 0 & 0 & 0 \\
 0 & 0 & 0 & 1 & 0 & 0 \\
 0 & 0 & 0 & 0 & 0 & 0 \\
 0 & 1 & 0 & 0 & 0 & 0 \\
 0 & 0 & 0 & 0 & 0 & 0 \\
 0 & 0 & 0 & 0 & 0 & 0
\end{array}
\right),
$$
\end{minipage}\qquad \qquad \qquad \qquad
\begin{minipage}{3cm}
\scriptsize
$$\lambda_{12}=
\left(
\begin{array}{cccccc}
 0 & 0 & 0 & 0 & 0 & 0 \\
 0 & 0 & 0 & -i & 0 & 0 \\
 0 & 0 & 0 & 0 & 0 & 0 \\
 0 & i & 0 & 0 & 0 & 0 \\
 0 & 0 & 0 & 0 & 0 & 0 \\
 0 & 0 & 0 & 0 & 0 & 0
\end{array}
\right),
$$
\end{minipage}\qquad \qquad \qquad
\\
\begin{minipage}{3cm}
\scriptsize
$$\lambda_{13}=
\left(
\begin{array}{cccccc}
 0 & 0 & 0 & 0 & 0 & 0 \\
 0 & 0 & 0 & 0 & 0 & 0 \\
 0 & 0 & 0 & 1 & 0 & 0 \\
 0 & 0 & 1 & 0 & 0 & 0 \\
 0 & 0 & 0 & 0 & 0 & 0 \\
 0 & 0 & 0 & 0 & 0 & 0
\end{array}
\right),
$$
\end{minipage}\qquad \qquad \qquad
\begin{minipage}{3cm}
\scriptsize
$$\lambda_{14}=
\left(
\begin{array}{cccccc}
 0 & 0 & 0 & 0 & 0 & 0 \\
 0 & 0 & 0 & 0 & 0 & 0 \\
 0 & 0 & 0 & -i & 0 & 0 \\
 0 & 0 & i & 0 & 0 & 0 \\
 0 & 0 & 0 & 0 & 0 & 0 \\
 0 & 0 & 0 & 0 & 0 & 0
\end{array}
\right),
$$
\end{minipage}\qquad \qquad \qquad
\begin{minipage}{3cm}
\scriptsize
$$\lambda_{15}=
\left(
\begin{array}{cccccc}
 \frac{1}{\sqrt{6}} & 0 & 0 & 0 & 0 & 0 \\
 0 & \frac{1}{\sqrt{6}} & 0 & 0 & 0 & 0 \\
 0 & 0 & \frac{1}{\sqrt{6}} & 0 & 0 & 0 \\
 0 & 0 & 0 & -\sqrt{\frac{3}{2}} & 0 & 0 \\
 0 & 0 & 0 & 0 & 0 & 0 \\
 0 & 0 & 0 & 0 & 0 & 0
\end{array}
\right),
$$
\end{minipage}\qquad \qquad \qquad
\\
\begin{minipage}{3cm}
\scriptsize
$$\lambda_{16}=
\left(
\begin{array}{cccccc}
 0 & 0 & 0 & 0 & 1 & 0 \\
 0 & 0 & 0 & 0 & 0 & 0 \\
 0 & 0 & 0 & 0 & 0 & 0 \\
 0 & 0 & 0 & 0 & 0 & 0 \\
 1 & 0 & 0 & 0 & 0 & 0 \\
 0 & 0 & 0 & 0 & 0 & 0
\end{array}
\right),
$$
\end{minipage}\qquad \qquad \qquad
\begin{minipage}{3cm}
\scriptsize
$$\lambda_{17}=
\left(
\begin{array}{cccccc}
 0 & 0 & 0 & 0 & -i & 0 \\
 0 & 0 & 0 & 0 & 0 & 0 \\
 0 & 0 & 0 & 0 & 0 & 0 \\
 0 & 0 & 0 & 0 & 0 & 0 \\
 i & 0 & 0 & 0 & 0 & 0 \\
 0 & 0 & 0 & 0 & 0 & 0
\end{array}
\right),
$$
\end{minipage}\qquad \qquad \qquad
\begin{minipage}{3cm}
\scriptsize
$$\lambda_{18}=
\left(
\begin{array}{cccccc}
 0 & 0 & 0 & 0 & 0 & 0 \\
 0 & 0 & 0 & 0 & 1 & 0 \\
 0 & 0 & 0 & 0 & 0 & 0 \\
 0 & 0 & 0 & 0 & 0 & 0 \\
 0 & 1 & 0 & 0 & 0 & 0 \\
 0 & 0 & 0 & 0 & 0 & 0
\end{array}
\right),
$$
\end{minipage}\qquad \qquad \qquad
\\
\begin{minipage}{3cm}
\scriptsize
$$\lambda_{19}=
\left(
\begin{array}{cccccc}
 0 & 0 & 0 & 0 & 0 & 0 \\
 0 & 0 & 0 & 0 & -i & 0 \\
 0 & 0 & 0 & 0 & 0 & 0 \\
 0 & 0 & 0 & 0 & 0 & 0 \\
 0 & i & 0 & 0 & 0 & 0 \\
 0 & 0 & 0 & 0 & 0 & 0
\end{array}
\right),
$$
\end{minipage}\qquad \qquad \qquad
\begin{minipage}{3cm}
\scriptsize
$$\lambda_{20}=
\left(
\begin{array}{cccccc}
 0 & 0 & 0 & 0 & 0 & 0 \\
 0 & 0 & 0 & 0 & 0 & 0 \\
 0 & 0 & 0 & 0 & 1 & 0 \\
 0 & 0 & 0 & 0 & 0 & 0 \\
 0 & 0 & 1 & 0 & 0 & 0 \\
 0 & 0 & 0 & 0 & 0 & 0
\end{array}
\right),
$$
\end{minipage}\qquad \qquad \qquad
\begin{minipage}{3cm}
\scriptsize
$$\lambda_{21}=
\left(
\begin{array}{cccccc}
 0 & 0 & 0 & 0 & 0 & 0 \\
 0 & 0 & 0 & 0 & 0 & 0 \\
 0 & 0 & 0 & 0 & -i & 0 \\
 0 & 0 & 0 & 0 & 0 & 0 \\
 0 & 0 & i & 0 & 0 & 0 \\
 0 & 0 & 0 & 0 & 0 & 0
\end{array}
\right),
$$
\end{minipage}\qquad \qquad \qquad
\\
\begin{minipage}{3cm}
\scriptsize
$$\lambda_{22}=\left(
\begin{array}{cccccc}
 0 & 0 & 0 & 0 & 0 & 0 \\
 0 & 0 & 0 & 0 & 0 & 0 \\
 0 & 0 & 0 & 0 & 0 & 0 \\
 0 & 0 & 0 & 0 & 1 & 0 \\
 0 & 0 & 0 & 1 & 0 & 0 \\
 0 & 0 & 0 & 0 & 0 & 0
\end{array}
\right),
$$
\end{minipage}\qquad \qquad \qquad
\begin{minipage}{3cm}
\scriptsize
$$\lambda_{23}=
\left(
\begin{array}{cccccc}
 0 & 0 & 0 & 0 & 0 & 0 \\
 0 & 0 & 0 & 0 & 0 & 0 \\
 0 & 0 & 0 & 0 & 0 & 0 \\
 0 & 0 & 0 & 0 & -i & 0 \\
 0 & 0 & 0 & i & 0 & 0 \\
 0 & 0 & 0 & 0 & 0 & 0
\end{array}
\right),
$$
\end{minipage}\qquad \qquad \qquad
\begin{minipage}{3cm}
\scriptsize
$$\lambda_{24}=
\left(
\begin{array}{cccccc}
 \frac{1}{\sqrt{10}} & 0 & 0 & 0 & 0 & 0 \\
 0 & \frac{1}{\sqrt{10}} & 0 & 0 & 0 & 0 \\
 0 & 0 & \frac{1}{\sqrt{10}} & 0 & 0 & 0 \\
 0 & 0 & 0 & \frac{1}{\sqrt{10}} & 0 & 0 \\
 0 & 0 & 0 & 0 & \frac{-4}{\sqrt{10}} & 0 \\
 0 & 0 & 0 & 0 & 0 & 0
\end{array}
\right),
$$
\end{minipage}\qquad \qquad \qquad
\end{table}

\newpage

\noindent
\begin{table}[!ht]
{{\bf{Table 1 Continued:  }}
\begin{center}
\section*{``Generalized" Gell-Mann and Flavor $\mathbf{U(1)}$ Singlet Matrices}
\end{center}
\doublerule
}

\begin{minipage}{3cm}
\scriptsize
$$\lambda_{25}=
\left(
\begin{array}{cccccc}
 0 & 0 & 0 & 0 & 0 & 1 \\
 0 & 0 & 0 & 0 & 0 & 0 \\
 0 & 0 & 0 & 0 & 0 & 0 \\
 0 & 0 & 0 & 0 & 0 & 0 \\
 0 & 0 & 0 & 0 & 0 & 0 \\
 1 & 0 & 0 & 0 & 0 & 0
\end{array}
\right),
$$
\end{minipage}\qquad \qquad \qquad
\begin{minipage}{3cm}
\scriptsize
$$\lambda_{26}=
\left(
\begin{array}{cccccc}
 0 & 0 & 0 & 0 & 0 & -i \\
 0 & 0 & 0 & 0 & 0 & 0 \\
 0 & 0 & 0 & 0 & 0 & 0 \\
 0 & 0 & 0 & 0 & 0 & 0 \\
 0 & 0 & 0 & 0 & 0 & 0 \\
 i & 0 & 0 & 0 & 0 & 0
\end{array}
\right),
$$
\end{minipage}\qquad \qquad \qquad
\begin{minipage}{3cm}
\scriptsize
$$\lambda_{27}=
\left(
\begin{array}{cccccc}
 0 & 0 & 0 & 0 & 0 & 0 \\
 0 & 0 & 0 & 0 & 0 & 1 \\
 0 & 0 & 0 & 0 & 0 & 0 \\
 0 & 0 & 0 & 0 & 0 & 0 \\
 0 & 0 & 0 & 0 & 0 & 0 \\
 0 & 1 & 0 & 0 & 0 & 0
\end{array}
\right),
$$
\end{minipage}
\\
\begin{minipage}{3cm}
\scriptsize
$$\lambda_{28}=
\left(
\begin{array}{cccccc}
 0 & 0 & 0 & 0 & 0 & 0 \\
 0 & 0 & 0 & 0 & 0 & -i \\
 0 & 0 & 0 & 0 & 0 & 0 \\
 0 & 0 & 0 & 0 & 0 & 0 \\
 0 & 0 & 0 & 0 & 0 & 0 \\
 0 & i & 0 & 0 & 0 & 0
\end{array}
\right),
$$
\end{minipage}\qquad \qquad \qquad
\begin{minipage}{3cm}
\scriptsize
$$\lambda_{29}=
\left(
\begin{array}{cccccc}
 0 & 0 & 0 & 0 & 0 & 0 \\
 0 & 0 & 0 & 0 & 0 & 0 \\
 0 & 0 & 0 & 0 & 0 & 1 \\
 0 & 0 & 0 & 0 & 0 & 0 \\
 0 & 0 & 0 & 0 & 0 & 0 \\
 0 & 0 & 1 & 0 & 0 & 0
\end{array}
\right),
$$
\end{minipage}\qquad \qquad \qquad
\begin{minipage}{3cm}
\scriptsize
$$\lambda_{30}=
\left(
\begin{array}{cccccc}
 0 & 0 & 0 & 0 & 0 & 0 \\
 0 & 0 & 0 & 0 & 0 & 0 \\
 0 & 0 & 0 & 0 & 0 & -i \\
 0 & 0 & 0 & 0 & 0 & 0 \\
 0 & 0 & 0 & 0 & 0 & 0 \\
 0 & 0 & i & 0 & 0 & 0
\end{array}
\right),
$$
\end{minipage}\qquad \qquad \qquad
\\
\begin{minipage}{3cm}
\scriptsize
$$\lambda_{31}=
\left(
\begin{array}{cccccc}
 0 & 0 & 0 & 0 & 0 & 0 \\
 0 & 0 & 0 & 0 & 0 & 0 \\
 0 & 0 & 0 & 0 & 0 & 0 \\
 0 & 0 & 0 & 0 & 0 & 1 \\
 0 & 0 & 0 & 0 & 0 & 0 \\
 0 & 0 & 0 & 1 & 0 & 0
\end{array}
\right),
$$
\end{minipage}\qquad \qquad \qquad
\begin{minipage}{3cm}
\scriptsize
$$\lambda_{32}=
\left(
\begin{array}{cccccc}
 0 & 0 & 0 & 0 & 0 & 0 \\
 0 & 0 & 0 & 0 & 0 & 0 \\
 0 & 0 & 0 & 0 & 0 & 0 \\
 0 & 0 & 0 & 0 & 0 & -i \\
 0 & 0 & 0 & 0 & 0 & 0 \\
 0 & 0 & 0 & i & 0 & 0
\end{array}
\right),
$$
\end{minipage}\qquad \qquad \qquad
\begin{minipage}{3cm}
\scriptsize
$$\lambda_{33}=
\left(
\begin{array}{cccccc}
 0 & 0 & 0 & 0 & 0 & 0 \\
 0 & 0 & 0 & 0 & 0 & 0 \\
 0 & 0 & 0 & 0 & 0 & 0 \\
 0 & 0 & 0 & 0 & 0 & 0 \\
 0 & 0 & 0 & 0 & 0 & 1 \\
 0 & 0 & 0 & 0 & 1 & 0
\end{array}
\right),
$$
\end{minipage}\qquad \qquad \qquad
\\
\begin{minipage}{3cm}
\scriptsize
$$\lambda_{34}=
\left(
\begin{array}{cccccc}
 0 & 0 & 0 & 0 & 0 & 0 \\
 0 & 0 & 0 & 0 & 0 & 0 \\
 0 & 0 & 0 & 0 & 0 & 0 \\
 0 & 0 & 0 & 0 & 0 & 0 \\
 0 & 0 & 0 & 0 & 0 & -i \\
 0 & 0 & 0 & 0 & i & 0
\end{array}
\right),
$$
\end{minipage}\qquad \qquad \qquad
\begin{minipage}{3cm}
\scriptsize
$$\lambda_{35}=
\left(
\begin{array}{cccccc}
 \frac{1}{\sqrt{15}} & 0 & 0 & 0 & 0 & 0 \\
 0 & \frac{1}{\sqrt{15}} & 0 & 0 & 0 & 0 \\
 0 & 0 & \frac{1}{\sqrt{15}} & 0 & 0 & 0 \\
 0 & 0 & 0 & \frac{1}{\sqrt{15}} & 0 & 0 \\
 0 & 0 & 0 & 0 & \frac{1}{\sqrt{15}} & 0 \\
 0 & 0 & 0 & 0 & 0 & \frac{-5}{\sqrt{15}}
\end{array}
\right),
$$
\end{minipage}\qquad \qquad \qquad \qquad \qquad
\begin{minipage}{3cm}
\scriptsize
$$\lambda_{0}=
\frac{1}{\sqrt{3}}\left(
\begin{array}{cccccc}
 1 & 0 & 0 & 0 & 0 & 0 \\
 0 & 1 & 0 & 0 & 0 & 0 \\
 0 & 0 & 1 & 0 & 0 & 0 \\
 0 & 0 & 0 & 1 & 0 & 0 \\
 0 & 0 & 0 & 0 & 1 & 0 \\
 0 & 0 & 0 & 0 & 0 & 1
\end{array}
\right).
$$
\end{minipage}
\end{table}

\begin{table}[!ht]
{\bf{Table 2:  }}
\begin{center}
\section*{Non-Zero Totally Anti-Symmetric $\mathbf{SU(6)}$ Structure Constants $\boldsymbol{f_{ijk}}$}
\end{center}
}
\doublerule
\begin{center}
\renewcommand{\arraystretch}{.85}
\begin{tabular}{lclclclc}
   \centering
$i\quad j\quad k$&$f_{ijk}$&$i\quad j\quad k$&$f_{ijk}$& $i\quad j\quad k$&$f_{ijk}$&$i\quad j\quad k$&$f_{ijk}$\\
\hline
        $1\quad   2\quad   3$   &   $ 1$    & $1\quad   4\quad   7$   &   $ 1/2$         & $1\quad   5\quad   6$   &   $ -1/2$       & $1\quad   9\quad   12$   &   $ 1/2$  \\
        $1\quad   10\quad   11$   &   $ -1/2$ & $1\quad   16\quad   19$   &   $ 1/2$         & $1\quad   17\quad   18$   &   $ -1/2$       & $1\quad   25\quad   28$   &   $ 1/2$  \\
        $1\quad   26\quad   27$   &   $ -1/2$ & $2\quad   4\quad   6$   &   $ 1/2$         & $2\quad   5\quad   7$   &   $ 1/2$        & $2\quad   9\quad   11$   &   $ 1/2$  \\
        $2\quad   10\quad   12$   &   $ 1/2$  & $2\quad   16\quad   18$   &   $ 1/2$         & $2\quad   17\quad   19$   &   $ 1/2$        & $2\quad   25\quad   27$   &   $ 1/2$  \\
        $2\quad   26\quad   28$   &   $ 1/2$  & $3\quad   4\quad   5$   &   $ 1/2$         & $3\quad   6\quad   7$   &   $ -1/2$       & $3\quad   9\quad   10$   &   $ 1/2$  \\
        $3\quad   11\quad   12$   &   $ -1/2$ & $3\quad   16\quad   17$   &   $ 1/2$         & $3\quad   18\quad   19$   &   $ -1/2$       & $3\quad   25\quad   26$   &   $ 1/2$  \\
        $3\quad   27\quad   28$   &   $ -1/2$ & $4\quad   5\quad   8$   &   $ \sqrt{3}/2$ & $4\quad   9\quad   14$   &   $ 1/2$        & $4\quad   10\quad   13$   &   $ -1/2$ \\
\end{tabular}
\end{center}
{\label{Table:structure}
\end{table}

\clearpage

{\bf{Table 3:  }}
\begin{center}
\section*{ Non-Zero $\bf{SU(6)}$ Commutators  }
\end{center}

\doublerule  

\footnotesize

\begin{multicols}{2}
\noindent\(1,[V_3, V_{\pi^+}]=\left(
%
\right)=(-2 \sqrt{\frac{2}{5}} V_{24}+2 \sqrt{\frac{3}{5}} V_{35})\)
\label{Table:3}
\end{multicols}

\newpage

\begin{table}[!ht]
{\bf{Table 4:  }}
\begin{center}
\section*{ Non-Zero Totally Symmetric $\mathbf{SU(6)}$ Structure Constant Related Tensors $\boldsymbol{d_{ijk}}$}
\end{center}
}

\doublerule
\begin{center}
\renewcommand{\arraystretch}{.85}
\begin{tabular}{lclclclc}
   \centering

$i\quad j\quad k$&$d_{ijk}$&$i\quad j\quad k$&$d_{ijk}$& $i\quad j\quad k$&$d_{ijk}$&$i\quad j\quad k$&$d_{ijk}$\\
\hline
        $0\quad j\quad k$&$\delta_{jk}/\sqrt{3}$         & $1\quad  1\quad  8$&$      1/\sqrt{3}$         & $1\quad  1\quad  15$&$      1/\sqrt{6}$&
        $1\quad  1\quad  24$&$      1/\sqrt{10}$   \\
              $1\quad  1\quad  35$&$      1/\sqrt{15}$        & $1\quad  4\quad  6$&$      1/2$&
        $1\quad  5\quad  7$&$      1/2$                & $1\quad  9\quad  11$&$      1/2$ \\                $1\quad  10\quad  12$&$      1/2$&
        $1\quad  16\quad  18$&$      1/2$                & $1\quad  17\quad  19$&$      1/2$                & $1\quad  25\quad  27$&$      1/2$                \\
        $1\quad  26\quad  28$&$      1/2$                & $2\quad  2\quad  8$&$      1/\sqrt{3}$         & $2\quad  2\quad  15$&$      1/\sqrt{6}$&
        $2\quad  2\quad  24$&$      1/\sqrt{10}$  \\       $2\quad  2\quad  35$&$      1/\sqrt{15}$        & $2\quad  4\quad  7$&$      -1/2$&
        $2\quad  5\quad  6$&$      1/2$                & $2\quad  9\quad  12$&$      -1/2$     \\           $2\quad  10\quad  11$&
        $      1/2$&
        $2\quad  16\quad  19$&$      -1/2$               & $2\quad  17\quad  18$&$      1/2$                & $2\quad  25\quad  28$&$      -1/2$               \\
        $2\quad  26\quad  27$&$      1/2$                & $3\quad  3\quad  8$&$      1/\sqrt{3}$         & $3\quad  3\quad  15$&$      1/\sqrt{6}$&
        $3\quad  3\quad  24$&$      1/\sqrt{10}$   \\
\end{tabular}
\end{center}
{\label{Table:4}
\end{table}

\begin{table}[!ht]
{{\bf{Table 4 Continued:  }}

\begin{center}
\section*{ Non-Zero Totally Symmetric $\mathbf{SU(6)}$ Structure Constants $\boldsymbol{d_{ijk}}$}
\end{center}
}

\doublerule

\begin{center}
\renewcommand{\arraystretch}{.85}
\begin{tabular}{lclclclc}
   \centering

$i\quad j\quad k$&$d_{ijk}$&$i\quad j\quad k$&$d_{ijk}$& $i\quad j\quad k$&$d_{ijk}$&$i\quad j\quad k$&$d_{ijk}$\\
\hline
$12\quad  12\quad  24$   & $   1/\sqrt{10}$         & $12\quad  12\quad  35$   & $   1/\sqrt{15}$         & $12\quad  18\quad  23$ & $   -1/2$                & $12\quad  19\quad  22$   & $   1/2$                 \\
        $12\quad  27\quad  32$   & $   -1/2$                & $12\quad  28\quad  31$   & $   1/2$                 & $13\quad  13\quad  15$ & $   -(1/\sqrt{6})$       & $13\quad  13\quad  24$   & $   1/\sqrt{10}$         \\
        $13\quad  13\quad  35$   & $   1/\sqrt{15}$         & $13\quad  20\quad  22$   & $   1/2$                 & $13\quad  21\quad  23$ & $   1/2$                 & $13\quad  29\quad  31$   & $   1/2$                 \\
        $13\quad  30\quad  32$   & $   1/2$                 & $14\quad  14\quad  15$   & $   -(1/\sqrt{6})$       & $14\quad  14\quad  24$ & $   1/\sqrt{10}$         & $14\quad  14\quad  35$   & $   1/\sqrt{15}$         \\
        $14\quad  20\quad  23$   & $   -1/2$                & $14\quad  21\quad  22$   & $   1/2$                 & $14\quad  29\quad  32$ & $   -1/2$                & $14\quad  30\quad  31$   & $   1/2$                 \\
        $15\quad  15\quad  15$   & $   -2/\sqrt{6}$          & $15\quad  15\quad  24$   & $   1/\sqrt{10}$         & $15\quad  15\quad  35$ & $   1/\sqrt{15}$         & $15\quad  16\quad  16$   & $   1/(2\sqrt{6})$       \\
        $15\quad  17\quad  17$   & $   1/(2\sqrt{6})$       & $15\quad  18\quad  18$   & $   1/(2\sqrt{6})$       & $15\quad  19\quad  19$ & $   1/(2\sqrt{6})$       & $15\quad  20\quad  20$   & $   1/(2\sqrt{6})$       \\
\end{tabular}
\end{center}
\end{table}
\clearpage

\newpage
{{\bf{\large Table 5  }}

\begin{center}
\section*{ Non-zero $\bf{SU(6)}$ Anti-Commutators}
\end{center}
\doublerule  

\footnotesize

\begin{multicols}{2}

\noindent\(1.\{V_3, V_{K^+}\}=\left(
%
\right)=
(\frac{V_{T_{b}^{+}}}{\sqrt{3}}\))

\end{multicols}
}

\newpage
{\small
\bibliographystyle{plain}

\begin{thebibliography}{99}

\bibitem{ODonnell:1981hgt}
  P.~J.~O'Donnell,
  \textit{Radiative decays of mesons},
  Rev.\ Mod.\ Phys.\  {\bf 53}, 673 (1981).
  doi:10.1103/RevModPhys.53.673

\bibitem{lie:1888groups}
 S.~Lie,
  \emph{Theorie der Transformationsgruppen, V.1, 1888},
  Reprinted by Chelsea Publishing Co., New York, N.Y. (1970),
  Modern Presentation and English Translation by Jo\"{e}l Merker,
  https://arxiv.org/pdf/1003.3202.pdf, (2010),  pp: (xii+638).

\bibitem{GellMann:1964tf}
  M.~Gell-Mann,
  \textit{The Symmetry group of vector and axial vector currents},
  Physics {\bf 1}, 63-75 (1964).


\bibitem{Cartan:1933thesis}
  E.~Cartan,
  \textit{Sur la structure des groupes de transformations finis et continus}, Thesis, Paris, 1894.


\bibitem{Tanabashi:2018xqp}
  M.~Tanabashi {\it et al.} [Particle Data Group],
  \textit{Review of Particle Physics},
  Phys.\ Rev.\ D {\bf 98}, 030001 (2018).

\bibitem{Slaughter:charge conjugate states}
  Particle charge conjugate states are utilized in this paper.

\bibitem{Dynkin1950AMS}
E.~B.~Dynkin,
\textit{The structure of semi-simple algebras},
Amer.\ Math.\ Soc.\ Translation 1950,
no. {\bf 17}, pp.143, (1950).

\bibitem{Racah:1961sj}
  G.~Racah,
  \textit{Group theory and spectroscopy},
  CERN-61-08, 6 March 1961.

\bibitem{behrends1962simple}
  R.~E.~Behrends, J.~Dreitlein, C.~Fronsdal, and W.~Lee,
  \textit{Simple groups and strong interaction symmetries},
  Rev.\ Mod.\ Phys.  {\bf 34}, 1, (1962).

\bibitem{Lichtenberg:1978pc}
  D.~B.~Lichtenberg,
  \textit{ Unitary Symmetry And Elementary Particles, Second Edition} (Academic Press, New York, 1978).

\bibitem{Slansky:1981yr}
  R.~Slansky,
  \textit{Group Theory for Unified Model Building},
  Phys.\ Rept.\  {\bf 79}, 1 (1981).
  doi:10.1016/0370-1573(81)90092-2

\bibitem{georgi:1999lie}
 Howard~M.~Georgi,
 \textit{Lie Algebras in Particle Physics: From Isospin to Unified Theories, 2nd ed.},
 Westview Press:Perseus, (1999), 331 p.

\bibitem{cahn:1984lie}
 Robert~N.~Cahn,
 \textit{Semi-Simple Lie Algebras and Their Representations},
  The Benjamin/Cummings Publishing Company,
  Advanced Book Program, Menlo Park, California, (1984).

\bibitem{Hamermesh:1123140}
 Morton~Hamermesh,
  \textit{Group theory and its application to physical problems},
  Dover edition, first published in 1989, is an
  unabridged, corrected republication of the second
  (corrected) printing (1964) of the work first published by
  Addison-Wesley Publishing Company, Inc., Reading,
  Massachusetts, 1962, in its "Addison-Wesley Series in Physics".

\bibitem{Oneda:1985wf}
  S.~Oneda and K.~Terasaki,
  \textit{Algebraic Approach In Quantum Chromodynamics And Electroweak Theory},
  Prog.\ Theor.\ Phys.\ Suppl.\  {\bf 82}, 1 (1985).


\bibitem{Oneda:1991wz}
  S.~Oneda and Y.~Koide,
  \textit{Asymptotic symmetry and its implication in elementary particle physics}, World Scientific Pub. Co., (1991).

\bibitem{jones:1998groups}
 H.~F.~Jones,
  \textit{Groups, Representations and Physics, 2nd ed.},
  Taylor and Francis Group, (1998).

\bibitem{Shapiro:2017lectures}
  J.~A.~Shapiro,
  \textit{Group Theory in Physics, lecture notes, Rutgers University},
  New Brunswick, NJ, 2017.

\bibitem{Weyl:2016Stanford}
  Bell, John L. and Korté, Herbert, \\
  Article on Hermann~Weyl,
  \textit{The Stanford Encyclopedia of Philosophy (Winter 2016 Edition),} Principal Editor: Edward~N.~Zalta\\
  https://plato.stanford.edu/archives/win2016/entries/weyl/ \\
  Publisher: Metaphysics Research Lab, Stanford University, Stanford, CA, USA.

\bibitem{Weyl:1927vd}
 Hermann~Weyl,
  \textit{Quantum mechanics and group theory},
  Z.\ Phys.\  {\bf 46}, 1 (1927).
  doi:10.1007/BF02055756

\bibitem{Christoph:2010}
  Christoph~L{\"u}deling,
  \textit{Group Theory (for Physicists)},\\
  http://www.th.physik.uni-bonn.de/nilles/people/luedeling/grouptheory/data/grouptheorynotes.pdf



\bibitem{Cox:1973}
  H.~S.~M.~Coxeter,
\textit{Regular Polytopes, 3rd ed.},
 Dover Publications, Inc., New York, 1973.

\bibitem{Slaughter:2011xs}
  M.~D.~Slaughter,
 \textit{Magnetic moments of the ground-state $J^P=(3/2)^{+}$ baryon decuplet},
  Phys.\ Rev.\ D {\bf 84}, 071303 (2011),
  doi:10.1103, \ Phys.\ Rev.\ D {\bf 84}, 071303,
  [arXiv:1107.2838 [hep-ph]].

\bibitem{Slaughter:2008zd}
  M.~D.~Slaughter,
  \textit{Gamma N Delta Form Factors and Wigner Rotations},
  Phys.\ Rev.\  C {\bf 80}, 038201 (2009).


\bibitem{Oneda:1989jr}
  S.~Oneda, M.~D.~Slaughter and T.~Teshima,
  \textit{Hadron '89. Proceedings, 3rd International Conference on Hadron Spectroscopy, Ajaccio, France, September 23-27, 1989}.  Editors: F.~G.~Binon, J.~M.~Frere and J.~P.~Peigneux, (Gif-sur-Yvette, France: Ed. Frontieres (1989) 767 p), pp. 423-428.  See also: \textit{An Algebraic Approach To The Axial Vector, Vector, And Pseudoscalar Mesons: Radiative Decays, Mass Relations, And The Singlet Electromagnetic Current},
  Maryland Univ. College Park - PP 90-075 (89, rec. Dec.) 7 p.


\bibitem{Oneda:1989ik}
  S.~Oneda, K.~Terasaki and M.~D.~Slaughter,
  \textit{On Broken Flavor Symmetry in Particle Phys}ics,
  KEK Library Report No. MDDP-PP-89-130, University of Maryland, College Park, 1989.
 [http://ccdb4fs.kek.jp/cgi-bin/img/allpdf?198904122].

  \bibitem{Slaughter:1988hx}
  M.~D.~Slaughter and S.~Oneda,
  $1^{--}$ \textit{to} $0^{-+}$ \textit{Meson Radiative And Pionic Transitions And Mass Splittings},
  Phys.\ Rev.\  D {\bf 39}, 2062 (1989).


\bibitem{Oneda:1979mp}
  S.~Oneda, T.~Tanuma, M.~D.~Slaughter,
  \textit{Why Are The Isoscalar Neutral Current Axial - Vector Couplings And Isoscalar Nucleon Anomalous Moments Small?},
  Phys.\ Lett.\  {\bf B88}, 343 (1979).

\bibitem{SlaughterOneda:1977PRL}
   Milton~D.~Slaughter and S.~Oneda, \textit{Intermultiplet Mixing of the Vector Mesons in a Nonperturbative Approach to Broken SU(4)}, Phys.\ Rev.\  Lett. {\bf 39}, 309 (1977). Erratum:Phys.\ Rev.\  Lett. {\bf 39}, 676 (1977).

\bibitem{YasueOnedaSlaughter:1983PRL}
   M.~Yasu$\grave{e}$, S.~Oneda, Milton~D.~Slaughter, \textit{Charge commutation relations, asymptotic $SU(2)_{L}$ symmetry, and the mass of the second Z boson in electroweak gauge theories}, Phys.\ Rev.\  D {\bf 30}, 174 (1984).

\bibitem{SlaughterOneda:1987PRL}
Milton~Dean~Slaughter and S.~Oneda, \textit{Theoretical Limit on $\iota\rightarrow \rho \gamma$ and Constraints on Glueball —$q\bar{q}$-Meson Couplings with the Pion and the Isovector Photon}, Phys. \ Rev. \ Lett. {\bf 59}, 1641 (1987).



\bibitem{Perl:1974}
  Martin~L.~Perl, \textit{High Energy Hadron Physics},
  (John Wiley and Sons, Inc., New York 1974).



\bibitem{Carruthers:1971}
  Peter~A.~Carruthers, \textit{Spin and Isospin in Particle Physics},
  (Gordon and Breach, Science Publishers, Inc., New York 1971).

\bibitem{alfaro63}
   V. De Alfaro, S. Fubini, G. Furlan, and C. Rossetti,
  {\it {Currents in Hadron Physics}} (North-Holland, Amsterdam, 1973).


\bibitem{adler}S. L. Adler, \textit{Sum Rules for the Axial-Vector Coupling-Constant Renormalization in
$\beta$ Decay},
 Phys. Rev. {\bf {140}}, B736 (1965).

\bibitem{Weisberger}William I. Weisberger, \textit{Unsubtracted Dispersion Relations and the Renormalization of the Weak Axial-Vector Coupling Constants}, Phys. Rev. {\bf {143}}, 1302 (1966).

\bibitem{weinberg:vol2}Steven~Weinberg, \textit{The Quantum Theory of Fields,Volume II
Modern Applications} (Cambridge University Press, 1996).

\bibitem{Alvarez-Gaume:2013CERN}L. Álvarez-Gaumé and M. A. Vázquez-Mozo, \textit{Introductory Lectures on Quantum Field Theory}, [arXiv:hep-th/0510040v4 20 Feb 2013]

\bibitem{AlvarezGaume:1985ex}
  L.~Álvarez-Gaumé,
  \textit{An Introduction To Anomalies},
  NATO Sci.\ Ser.\ B {\bf 141} (1986).

\bibitem{Bilal:2008qx}
  A.~Bilal,
 \textit{Lectures on Anomalies},
  arXiv:0802.0634 [hep-th].



\end{thebibliography}

} 
\end{document}